\documentstyle[12pt,epsf]{article}
\begin{document}
\thispagestyle{plain}
\newcommand{\vk}{{\vec k}}
\newcommand{\vK}{{\vec K}}
\newcommand{\vb}{{\vec b}}
\newcommand{\vp}{{\vec p}}
\newcommand{\vq}{{\vec q}}
\newcommand{\vQ}{{\vec Q}}
\newcommand{\vx}{{\vec x}}
\newcommand{\tr}{{{\rm Tr}}}

\begin{center}

\vspace*{1.0cm}

\begin{flushleft}
\end{flushleft}

\Large{\bf{Jet  Quenching 
in \\Thin Quark-Gluon Plasmas I: Formalism}}

\vspace{2.0cm}

\large{\bf{M. Gyulassy$^a$, {\rm P. L\'evai}$^{a,b}$, I. Vitev$^a$ }}

\vspace{1.0cm}

{\em ${}^a$ Department of Physics, Columbia University, \\ 
New York, NY, 10027, USA \\  
${}^b$ KFKI Research Institute for Particle and Nuclear Physics, \\ 
PO Box 49,  Budapest, 1525, Hungary  }\\[5ex]

{\large July 22, 1999}

\end{center}
\vspace{1.5cm}

\begin{abstract}

The modification and amplification of the gluon angular distribution produced
along with hard jets in nuclear collisions is computed.  We consider the
limit of a thin quark-gluon plasma, where the number of rescatterings of the
jet and gluons is small.  The focus is on jet quenching associated with
the formation of highly off-shell partons in hard scattering events involving
nuclei. The interference between the initial hard radiation amplitude, the
multiple induced Gunion-Bertsch radiation amplitudes, and gluon rescattering
amplitudes leads to an angular distribution that differs considerably
from both the standard DGLAP evolution and from the classical limit parton
cascading. The cases of a single and double rescattering are considered in 
detail, and a systematic method to compute all matrix elements for
the general case is developed.  A simple power law scaling of the angular
distribution with increasing number of rescatterings is found and used 
for  estimates of the fractional energy loss as a function
of the plasma thickness.

\end{abstract}

\newpage

\section{Introduction}

One of the new predicted  observables in nuclear collisions at RHIC
energies $(\surd s \sim 200$ AGeV) is jet
quenching~\cite{BJ}-\cite{MGXW92}.  While collisional energy loss of a
hard jet propagating through quark-gluon plasma has been estimated to
be modest~\cite{BJ,THOMA}\/: $dE_{coll}/dx \ll 1$ GeV/fm, for
accessible plasma temperatures, the induced radiative energy loss,
as estimated~\cite{MGMP,GPTW} using  the Gunion-Bertsch
~\cite{GUNION} non-abelian generalization
of the Bethe-Heitler formula for QED
is expected to be significantly larger with
$dE_{rad}/dx> {\rm few}$  GeV/fm.  An uncertainty principle analysis
in~\cite{BRODS} leads to a simple estimate
\begin{equation}
\frac{dE}{dx}\sim  \frac{\langle k_\perp^2\rangle}{2}
\; , \label{bh}
\end{equation} 
where $\langle k_\perp^2\rangle$ is the mean transverse momentum of
the radiated gluons.  In Ref.~\cite{MGXW} a model for non-abelian
energy loss was developed and a class of multiple scattering
diagrams lead to  $dE/dx \sim 1$ GeV/fm
independent of the path length.  However, in BDMPS~\cite{BDMPS} 
 gluon rescattering diagrams in the medium were shown
to induce a nonlinear enhancement of the energy loss because
 $ \langle k_\perp^2 \rangle =\mu^2 L/\lambda$.
The result for an incident very high energy jet penetrating a plasma of
thickness, $L$, in which the mean free path, $\lambda=1/(\sigma\rho)$,
and the color electric fields are screened on a scale $\mu$, 
is ~\cite{BDMPS} 
\begin{equation}
\frac{dE}{dx}\approx \alpha_s  {\langle p_\perp^2\rangle}
\propto \alpha_s \mu^2 \frac{L}{\lambda} \; . \label{BDPS}
\end{equation} 
Here $ {\langle p_\perp^2\rangle}$ is the jet broadening due to
multiple elastic scatterings of the jet in matter,
 and a factor of order unity
is suppressed.  
This estimate is derived for the situation
when  the radiation formation
length is large compared to the target thickness and the target is thick
compared to the mean free path:
$$\Delta L_f=E/\mu^2 \gg L\gg \lambda \; .$$ 
 In the other theoretical limit corresponding to 
Landau--Pomeranchuk--Migdal (LPM) ~\cite{LPM} regime in QED, 
 $L\gg \Delta L_f\gg \lambda$,  and 
the energy loss was shown to saturate as 
 ${dE}/{dx}\propto \alpha_s \sqrt{\rho E}$.

 The remarkable prediction that  the non-abelian $dE/dx$ increases linearly
 with the plasma thickness in nuclear collisions, suggests that the
 total jet energy loss could increase quadratically with the nuclear
 size.  However, several practical problems 
limit the application of the above results directly
to observable jet quenching in  nuclear collisions.
First, jets are produced in such reactions
with high initial virtualities in
hard pQCD processes, and thus induced radiation from
secondary interactions must compete with the usual  broad gluon shower
accompanying a hard pQCD event. 

Second, the observable energy loss depends on the detailed angular
 distribution of the gluons.  Unlike in electrodynamics where the
 energy of the final electron and photon can be directly measured, the
 energy of a single parton is only measurable when it is
 kinematically well separated from other jets. This is because the
 observable hadron showers have wide angular distributions.  Nearly
 collinear partons in particular are not resolvable because they are
 absorbed into the coherent hadronic wavefunctions.  Therefore, in
 order to test experimentally the non-abelian energy loss of jets,
 $dN_g/dyd^2 \vk_\perp$ of the induced gluon radiation must be known.
 The angular integrated energy loss, $dE/dx$, is insufficient to
 predict its experimental consequences.
 
Third, for finite nuclei  $L/\lambda<10$ is not very large, and
a large fraction  of the jets are
produced in the nuclear ``corona'' with $L/\lambda \le 3 $.
Thus the form of the angular distribution for only a few collisions
is needed. 

Fourth, a problem specific to nuclear collisions is that
 the very large transverse energy background
 due to minijets (the partons from which the dense quark-gluon plasma is made ) prevents the application
 of traditional transverse energy jet
 cone measurements of jets \cite{Dong}.
 Therefore, jet quenching must be inferred from the systematics
of leading hadrons at  moderate $p_\perp
 \sim 3-10$ GeV relative 
 to well know  spectra  in elementary nucleon-nucleon
 reactions. This step requires the interface of the parton level
shower calculations with hadronization phenomenology.

This paper is part I of a study of jet quenching 
concentrating on the formal problem of computing
the angular distribution gluon  radiation from
off-shell jets produced in a dense but thin medium.
 In the subsequent paper II~\cite{GLVII},
 we  encode  the results of this paper 
 into a nuclear collision 
 event generator (HIJING1.36~\cite{hijing}
using the Lund JETSET~~\cite{JETSET} string
 fragmentation scheme for hadronization) and 
 compute the hadron inclusive
jet quenching pattern.
Note that in a separate study~\cite{WiedGY},
the transverse momentum dependence of the LPM
effect in QED was discussed in another formalism
and compared to the analytic  results for QCD derived
in this paper.

 Multiple interactions in QCD are complicated by the color charge of
 the radiated gluons. This leads to interference between the
 production amplitudes and the rescattering amplitudes that makes it
 impossible to factor the cross section into a product of the elastic
 cross section and the gluon distribution.  In BDMPS, a great
 technical simplification was achieved by considering only the angular
 integrated $dN_g/d\omega$, energy distribution.  The freedom to shift
 the transverse momentum, $\vk_\perp$, integrations was used often to
 reduce the multiple collision problem into tractable form.  In
 addition, a large $n_s=L/\lambda\gg 1 $ and $N_c\gg 1$ approximations
 were employed to convert the problem into an elegant equivalent
 Schroedinger equation problem valid under the conditions that
$$1/\mu\ll \lambda\ll L \ll L_f=E/\mu^2 \; .$$ 
An alternative path integral approach  proposed
in~\cite{Zakharov} obtained 
similar results (see also \cite{WiedGY}).

Unfortunately, the continuum  approximations developed
in \cite{BDMS} are not applicable to thin
plasmas and more importantly
they do not apply at all to the problem of computing the angular
distributions. The main complication is that the angular distributions
require a complete treatment of all gluon and jet final state
interactions.  In this paper, we simply follow the direct albeit tedious
route of computing the sum of squares of all diagrams with
finite number of rescatterings.
In the thin plasma limit, the  main simplicity arises from
the fact that for $n_s \le 3$ the number of
amplitudes $2^{n_s+1}-1$ is  still small enough
for practical calculations.

Historically, the on-shell single rescattering case, $n_s=1$, was
first considered by Gunion and Bertsch~\cite{GUNION} (GB).  They
estimated the hadron rapidity distribution due to gluon bremsstrahlung
associated with a single elastic valence quark scattering in the
Low-Nussinov model.  The important difference compared to the
problem addressed in this paper is that GB computed the non-abelian
analog of the Bethe-Heitler formula while we are interested in the
analog of radiative energy loss in sudden processes similar to 
the beta decay.  The GB problem is to compute the soft radiation
associated with a single scattering of an incident on-shell quark
prepared in the remote past, i.e., $t_0=-\infty$, relative to the
collision time, $t_1$.
In our case, the radiation associated with a hard jet processes in
nuclear reactions must take into account the fact that the jet parton
rescatters within a short time , $t_1-t_0 \sim R_A$, after it is
produced.  The sudden appearance of the jet color dipole moment within
a time interval $\sim 1/E$ results in a broad angular distribution of
gluons even with no final state scattering.  The induced radiation
cased by rescattering necessarily interferes strongly with this hard
radiation as indicated in~\cite{GLVQM99}.

To model a thin plasma, we employ the static color-screened potential
model as considered in Gyulassy--Wang (GW) \cite{MGXW}.  The scattering
potential has  Fourier components
\begin{eqnarray}
V_i &=& V(\vec{q}_i\;) e^{-i \vq_i\cdot\vx_i}
 \; T_a(R)\otimes T_a(i)\; ,
\;\;
V(\vec{q}_i\;)=\frac{4 \pi \alpha_s}{\vec{q}_i^{\;2}+\mu^2} 
\; . \label{pot}
\end{eqnarray}
Here $\tr( T_a(R) T_b(R)) =\delta_{ab} C_R D_R/D_A$ is proportional
to the second order Casimir of the $D_R$ dimensional 
representation $R$ of the jet parton, and $D_A=N_c^2-1$.
In principle, the $i^{th}$ target could be in the $D_i$ dimensional
representation with Casimir
$C_{2}(i)$ so that $\tr( T_a(i) T_b(j))
 =\delta_{ab} \delta_{ij}C_2(i) D_i/D_A$. 
Note that in eq.~(\ref{pot}) $V_i$ 
corresponds to $A_i(\vq\,)/(-2i P^0)$ in the notation of Ref.~\cite{MGXW}.

For a thermally equilibrated medium at temperature $T$, the color screening
mass in pQCD is given by $\mu=4 \pi \alpha_s T^2$.  In addition, as in ordinary
plasmas, no gluon modes propagate below the plasma frequency, $\omega_{pl}\sim
\mu/\surd 3 $.  In practice, lattice QCD calculations of $\mu$ indicate
sizable nonperturbative corrections to the pQCD estimates for achievable
temperatures. Therefore, we simply take here $\mu\sim \omega_{pl}\sim 0.5$ GeV
as a characteristic infrared scale of the medium.
In perturbation theory, there is a  relation between the screening scale $\mu$ 
and the mean free path $\lambda$,
\begin{equation}
\mu^2/\lambda \approx 4\pi \alpha_s^2 \rho \; 
\end{equation}
where $\rho$ is the density of plasma partons weighed by appropriate color
factors. Another important scale for our problem is the Bethe-Heitler 
frequency,
\begin{equation}
\omega_{BH} \equiv \frac{1}{2}\mu^2 \lambda \gg \mu 
\end{equation}
in the dilute plasma approximation assumed here.

This paper is organized as follows: in Sec.~2 we derive the spectrum of the
gluons, radiated as a result of the sudden production of a high energy parton
in a highly localized wave-packet $J(p)$ in free space. The large radiative
energy loss, which we call ``self-quenching'', grows as $\Delta E \sim E_0
\log(E_0/\mu)$. The strong destructive interference of this initial amplitude
with the subsequent induced radiative and cascade amplitudes is considered in
Sec.~3.  The main formal results of this paper are derived in that section.
In Sec.~3.1 jet acoplanarity is reviewed
for the case of multiple elastic scattering without radiative energy loss.
The formal structure of joint  jet plus gluon probability density including
multiple rescatterings is derived in Sec~3.2.
A systematic procedure for extracting all the relevant matrix elements in the eikonal
approach is developed using a simple binary encoding scheme 
in Sec~3.3 to help automate the computation of phase factors and the color algebra.
In Sec~3.4 the interference form factors and their dependence
on the detailed plasma geometry is derived. These form factors
control how close the exact result is to the extreme
factorization
or  the classical parton cascade limits.
In Sec.~3.5, the physical interpretation of the relative conditional
probabilities computed from the square of the sum of amplitudes 
and the wavefunction renormalization necessary to compute the gluon angular
distribution as a function of $n_s$ is  presented.
 Sec.~4 illustrates in detail 
the  single rescattering case. We rearrange the amplitudes
into Gunion-Bertsch and gluon rescattering terms that reduce in special
limits to the known cases.  An additional form factor associated with
the transverse profile of the target is derived that helps to identify
the physical interpretation of the rearranged terms.
In the simpler ``broad'' target case, those form factors drop out.
The induced gluon
bremsstrahlung (beyond self-quenching) 
is shown to be strongly suppressed relative to
the classical limit due to the destructive LPM effect in QCD
at small angles. However, the angular
enhancement at large angles due to gluon rescattering remains
a key feature. The more
involved case of two scatterings is treated in Sec.~5. The 
analytical result shows  a power law enhanced  angular broadening 
of the gluon distribution relative to the case of $n_s=1$.
This approximate power law scaling, $\propto (1+C_A/C_R)^{n_s}$,
is used in Sec.~6 to extract an approximate $n_s$ independent
reduced quantum modification factor of the gluon angular distribution.
We apply this scaling to compute the angular integrated gluon energy spectrum, 
$dI^{(n_s)}/d\omega$ and the jet fractional energy loss $ \Delta E^{(n_s)}/E$.
The numerical results show surprisingly an  approximate linear dependence of
those quantities  on the plasma thickness at least up
to $L=n_s\lambda\sim 3 $ fm. The jet energy loss  remains small because the
wavefunction renormalization factor also grows rapidly with $n_s$.
As emphasized in the Summary
these estimates are only lower bounds
 because multi-gluon emission is not treated in this paper. The generalization 
to multi-gluon effects is deferred to ref.\cite{GLVII}.
Details of the $n_s=2,3$ matrix elements and their rearrangement
into the physical hard, Gunion-Bertsch, and gluon cascading
components as well as the  relevant color factors  are recorded in
Appendix A and C to ease the flow of the presentation.
In Appendix B   we compare the $n_s=2$ case
to the approximation of BDMPS where some of the terms in the amplitude
were neglected.  In Appendix  D
we derive the curious color ``wheel diagram'' that arises in the $n_s=3$ case.

\section{Hard Radiation and Self-Quenching}

Consider first the amplitude for the production of a hard jet with
momentum $P_0$ localized initially near $x_0^\mu=(t_0,\vec{0})$. We
denote this color independent amplitude by
\begin{equation}
{\cal{M}}_{J} = J(P)e^{i P x_0} \; . 
\end{equation} 
The hard vertex is localized within a distance $\sim 1/P_0$,
and the amplitude, $J(P)$, is assumed to vary
slowly with $P$ on the infrared screening
scale $\mu$ of the medium.  This spectrum of hard jets is of course softened
due to gluon radiation to first order in $g_s$ as shown in Fig. 1.

\begin{center}
\vspace*{7.0cm}
\includegraphics{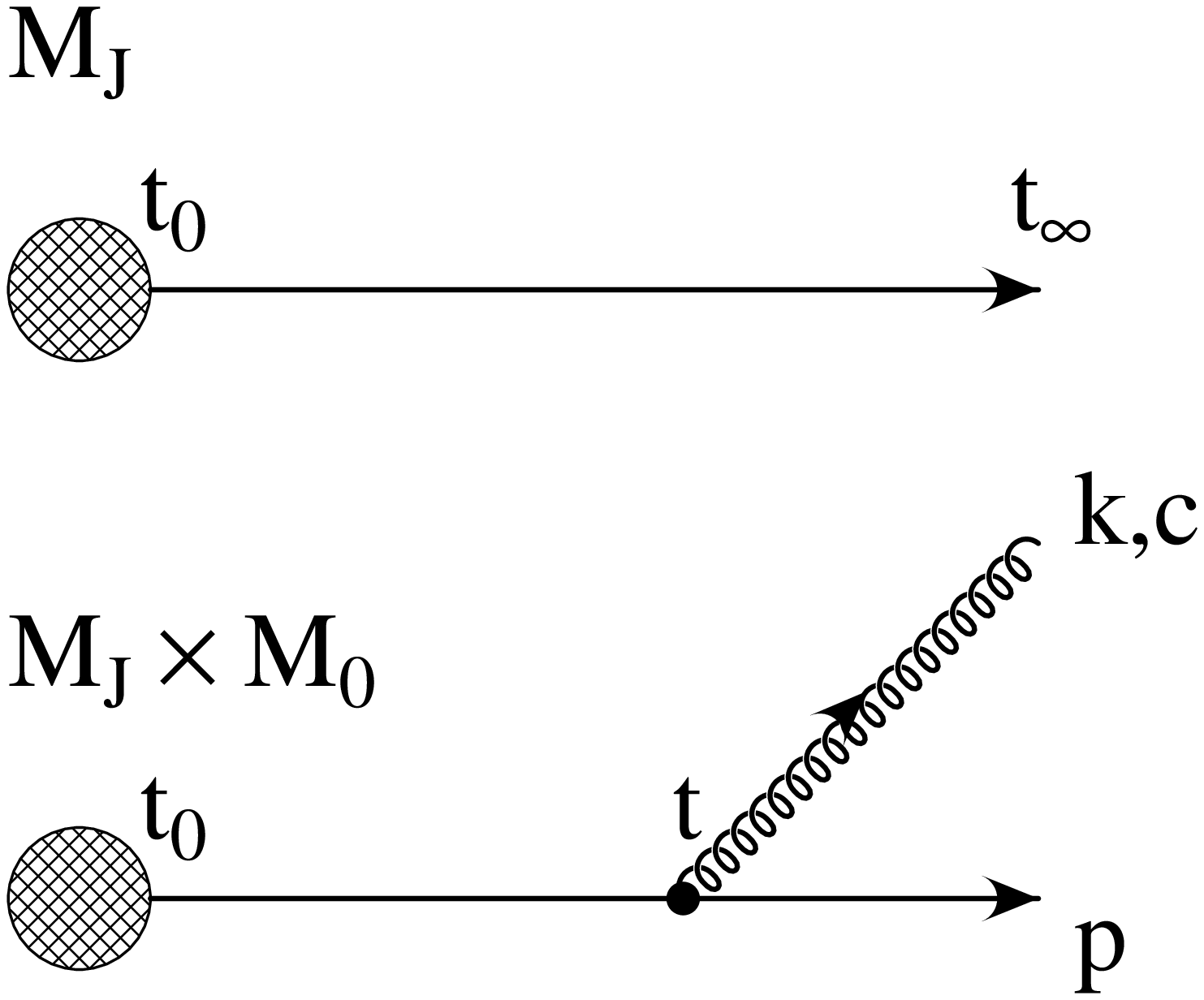}
\vskip -20pt
\begin{minipage}[t]{15.0cm}
{\small {\bf Fig.~1.}
The {\cal M}$_J$ hard jet production amplitude  and the 
{\cal M}$_J \otimes${\cal M}$_0$ soft gluon radiation amplitude.}
\end{minipage}
\end{center}
\vskip 4truemm

The final state in that case is characterized by a 
gluon and a jet with momenta
\begin{eqnarray}
k^\mu&=&(\omega,k_z,\vk_\perp)=[z
P_0^+,\frac{k_\perp^2}{z P_0^+},\vk_\perp]\; ,\nonumber \\
p^\mu&=&(p_0, p_z,\vec{Q}_\perp-\vk_\perp)=[(1-z)
P_0^+,\frac{(\vec{Q}_\perp-\vk_\perp)^2}{(1-z) P_0^+},
\vec{Q}_\perp-\vk_\perp]\; ,
\end{eqnarray}
where $(\cdots)$ denote their 4-momentum and $[\cdots]$ denote their
light-cone momentum coordinates. 

We emphasize that the ``z'' direction in this paper
refers to the unperturbed hard jet direction.
In nuclear reaction, the jets  are directed perpendicular
to the beam axis in the frame in which its 
longitudinal rapidity vanishes. 
Therefore $k_x,k_y,k_z$
actually correspond to a 90 degree rotation about
the $y$ axis of the  conventional
frame in which the z axis is parallel to the beam.
The initial ``transverse'' momentum $\vQ_\perp$ of the
jet thus accounts for both its longitudinal rapidity
as well as its azimuthal (symmetric) 
dependence about the beam. The energy loss, $\omega$, leads
to a suppression of the high $p_\perp$ distribution of the
fragments emerging from the jet.

The gluon polarization in the $A^+=0$
gauge is $\epsilon(k)=[0,(2 \vec{\epsilon}_\perp \cdot \vk_\perp)/(z
P_0^+), \vec{\epsilon}_\perp]$. We consider here 
eikonal kinematics where
\begin{eqnarray}
\omega&=&\sqrt{k_z^2+k_\perp^2 } \approx z E_0 +
\frac{k_\perp^2}{2 z E_0} \; ,\nonumber \\
p^0&=&\sqrt{p_z^2+p_\perp^2 }\approx (1-z) E_0 +
\frac{(\vec{Q}_\perp-\vk_\perp)^2}{2 (1-z) E_0} \; .
\end{eqnarray}

The amplitude to emit a soft gluon and the jet neglecting spin effects
can be expressed as
\begin{eqnarray}
{\cal{M}}_{J}\otimes{\cal{M}}_{0}&=& i \int d^4 P 
\; J(P) e^{i Px_0} \frac{(-i g_s T_c )\; \epsilon \cdot 
(2p+k)}{(p+k)^2} \delta^4(P-k-p)\approx \nonumber \\
&\approx & e^{i E_0 t_0} J(k+p) {\cal{M}}_{0}(k,p) \; ,
\label{self}\end{eqnarray}
where ${\cal{M}}_{0}$ is the radiation amplitude that in the eikonal
limit is given by
\begin{eqnarray}
{\cal{M}}_{0} &=&  - 2 i g_s {\vec{\epsilon}_\perp \cdot
\vk_\perp \over k^2_\perp } e^{i t_0 {k^2_\perp 
\over 2\omega}}  c \; .\label{m01}
\end{eqnarray}
Here, $c\equiv T_c(R)$ is the  $SU(N_c)$ color generator $T_c$ in the $D_R$
dimensional representation of  the jet parton with
$Tr(cc)=C_R D_R$.  

A simple direct way to recover the phase factor and the $1/k_\perp^2$
singularity of ${\cal M}_0$ is through the use time ordered
perturbation theory as emphasized in ~\cite{BDMPS}. This follows from
integrating over the emission time of the gluon
\begin{eqnarray}
\int_{t_0}^{\infty} dt \; e^{-\varepsilon |t|+i (E_f-E_i) t}
= {2\omega \over k^2_\perp } 
e^{i t_0 {k^2_\perp \over 2\omega}}   \; ,\label{intt}
\end{eqnarray}
where $$E_f-E_i= \omega+p_0 - E_0 \approx  k^2_\perp /2 \omega\; .$$
This method is particularly useful when final state interactions are
taken into account in the next sections.

The probability distribution  for the jet production is given by
\begin{equation}
d^{\, 3}{\cal D}^{(0)}=  D_R |{{\cal M}_J}|^2 
\frac{d^3{\vec p}}{(2\pi)^3 2p^0} 
\equiv \rho^{(0)}(\vp)
\frac{d^3{\vec p}}{(2\pi)^3 2|\vp\,|} \; ,
\label{jet0}
\end{equation}  
where $\rho^{(0)}(p)$ is proportional to the invariant
hard production probability density with no elastic rescatterings
of the jet nor radiative energy loss. The pQCD jet distribution
is approximately a power law $1/p_\perp^{2q}$ relative to the
beam axis and uniformly distributed in longitudinal
rapidity. The power $q\sim 3$ depending
on the beam energy and the $p_\perp$ range under consideration. 
However, in terms of our rotated variables with $p_z\leftrightarrow p_x$
$$ \rho^{(0)}(\vp\,)\sim c/(p_z^2+p_y^2)^q $$
independent of our $p_x$ that is parallel to the beam direction.
We emphasize this point because unlike the usual energy loss problem
of a beam passing through matter, the peculiar longitudinally
expanding cylindrical geometry
of the matter in nucleus collisions makes applications
of simple formulas in the rotated frame somewhat unintuitive.  
 
The double inclusive jet plus gluon probability distribution is given by
\begin{equation}
d^{\, 6}{\cal D}^{(0)}_{rad}= \rho^{(0)}_{rad}(k,p)
\frac{d^3 {\vec k}}{(2\pi)^3 2\omega} \frac{d^3 {\vec p}}{(2\pi)^3 2p^0}
\; ,
\end{equation}
where \begin{equation}
\rho^{(0)}_{rad}(k,p)\equiv \tr |{\cal{M}}_{J}\otimes{\cal{M}}_{0}|^2 
\end{equation}
is the double inclusive probability density for producing a jet and a gluon
in the case of no rescatterings.

As seen from (\ref{self},\ref{m01}) the double inclusive
density factorizes for soft gluons into a production probability
of an off-shell jet with $k+p$ and a 
distribution of gluons.
\begin{equation}
\omega \frac{d^{\, 3} N_{g}^{(0)}}{
d^3 {\vec k}} \approx  \frac{C_R \alpha_s}{\pi^2}\frac{1}
{k_\perp^2}\equiv R^{(0)}_g(\vk) 
\; ,
\label{rhog0}\end{equation}
where the dependence on $p$ is only implicit through the kinematic boundaries.
As in QED the radiation pattern associated with a hard jet
corresponds to a uniform rapidity distribution along the jet axis
with a broad  transverse transverse momentum distribution that is
uniform in $\log k_\perp^2$.

If the off-shell dependence
of the source amplitude is weak and the gluon is soft in the sense
$$\frac{1}{|\vp\,|} |J(|\vp\,|+\omega,\vp+\vk\,)|^2
\approx  \frac{1}{|\vp+\vk|} |J(|\vp+\vk|,\vp+\vk)|^2
\; \; ,$$
then we can write
\begin{equation}
\rho^{(0)}_{rad}(\vk,\vp)\approx 2(2\pi)^3
\rho^{(0)}(\vp+\vk)  R^{(0)}(\vk)
 \; .
\label{dn0}\end{equation}

The  gluon radiation associated with any hard pQCD process
leads to what one can think of as ``self-quenching'' of the jet in
the sense that the initially produced high energy jet parton loses a
substantial fraction of its energy
\begin{equation}
\Delta E_0 = C_R \frac{\alpha_s}{\pi} \int_0^1 
\frac{dz}{z} \int_{\mu^2}^{E^2} \frac{dk_\perp^2}{k_\perp^2} 
\; z E_0 \sim C_R \frac{2\alpha_s}{\pi} E_0 \;\log(E_0/\mu) \; .
\end{equation}
This occurs even in free space where there are no final state
 interactions.  In practice, the large logarithm implies that multiple
 gluon emission must also be considered. This leads to a Sudakov form
 factor for the jet and a probabilistic Alterelli-Parisi parton shower
 ~\cite{Field}. The final state multigluon shower can be calculated
 most readily by one of the many Monte--Carlo event generators such as
 PYTHIA~\cite{PYTHIA}.  Those generators encode empirical ``parton to
 hadron'' fragmentation functions and thus allow the detailed study of
 the effect of parton showering on the final hadron distributions. 
The magnitude of the
 self-quenching effect, i.e., initial and final state radiation, is
 found to be significant and is essential to describe accurately the
 high $p_\perp$ data as shown in~\cite{HIJINGPP}. Our main interest
 here is to derive the extra  quenching associated with gluon
 radiation induced by final state interactions of
the jet in a thin plasma.

\section{Gluon Radiation with Rescatterings}

In this section, the self-quenching case is generalized
to take into account 
possible multiple scattering of the jet and
gluon following the hard jet production. 

\subsection{Multiple Elastic Scattering}

The amplitude that jet undergoes $n_s$ elastic scatterings with  
momentum transfers $q_i=(0,q_{iz},\vq_{i \perp})$
is given by~\cite{MGXW}
\begin{eqnarray}
{\cal{M}}^{(n_s)}_{J}&=& \int d^4 P \; J(P) e^{i Px_0} 
  \int \prod_{i=1}^{n_s} \left ( \frac{d^3 \vq_i}{(2\pi)^3}
 \; (2P^0 T_{a_i}(i)\Delta(P+Q_{i-1})V(\vq_i) 
e^{-i \vq_i \vx_i}  \right) 
\nonumber \\[1.5ex]
 &\;& \times \; \delta^4(P-p+\sum q_i) \left (
T_{a_1}(R) \cdots T_{a_{n_s}}(R) \right ) \;  .
\end{eqnarray}
where $\Delta(p)=(p^2-m^2+i\epsilon)^{-1}$ is the jet propagator neglecting
spin effects, and the cumulative momentum transfer
$Q_i\equiv \sum_{j=1}^{i} q_j$. 
The invariant jet probability distribution together with
$n_s$ unobserved recoil partons in the target
and no gluon radiation is to lowest order (without Sudakov factor)  given by
\begin{equation}
d^{\, 3}{\cal D}^{(n_s)}_{el}= {  \left \langle \begin{array}{c}
\tr |{{\cal M}^{(n_s)}_J}|^2  \end{array}\right 
\rangle_{\vec x} } \frac{d^3{\vec p}}{(2\pi)^3 2p^0} 
=\rho^{(n_s)}_{el}(p)\frac{d^3{\vec p}}{(2\pi)^3 2p^0} 
\; . \label{elden}
\end{equation}  
The ensemble average over the location 
of  the scattering centers is taken 
as in~\cite{MGXW}
\begin{equation}
\langle \; \cdots \; \rangle_{\vec x}= \int \prod_{i=1}^{n_s}  
\left (d \vec{x}_{i \perp}  T(\vec{x}_{i \perp}) \; 
\int_0^\infty 
\frac{d(z_i-z_{i-1})}{\lambda_i} e^{-\frac{z_{i}-z_{i-1}}{\lambda_i}}
\theta(z_i-z_{i-1}) 
\right ) (\; \cdots \; )  \; .\label{zdis}
\end{equation}
In a static medium,
the average longitudinal separation between the centers is specified by the 
mean free path, $\lambda=\lambda_i$. An expanding medium,
on the other hand, has $\lambda_i>\lambda_{i-1}$ can only be approximately
treated with a constant mean free path.
 The transverse coordinates are distributed according
to a  normalized overlap function, $T(\vec{x}_\perp)\sim 1/(\pi R_\perp^2) $.
We assume that the transverse size $R_\perp$ of the target (relative to the jet
axis) is large
compared to the range of the interactions, $1/\mu$.

In the high energy ($p^0=E\gg\omega \gg \mu$) 
eikonal limit, going back to the picture including the momenta of the jet,
the integrations over $p_{iz}$ are fixed by residues approximately
to  $p_{iz} \approx E-p_{i\perp}^2/(2 E)$. The contributions from the
poles of the static potential are suppressed by a factor $\propto
exp(-\mu\lambda)$. In this model we neglect
the small collisional energy loss associated with target:
$dE_{el}/dx\approx 4 \alpha_s^2 T^2\log E/\mu\sim 0.3 $ GeV/fm~\cite{BJ,THOMA}.

Elastic scattering modifies the jet distribution  in
eq.~(\ref{elden}) in this eikonal limit as 
\begin{eqnarray}
\rho^{(n_s)}_{el}(p)&\approx& 
 \int \prod\limits_{i=1}^{n_s}  
\left (\frac{d^2 \vq_i}{(2\pi)^2} \, 
\frac{d^2 {\vq_i}^{\; \prime}}{(2\pi)^2} \; V(\vq_i)
V^*({\vq_i}^{\; \prime}) \frac{C_R C_2(i)}{D_A}
T(\vq_{i \perp}-{\vq_{i \perp}}^{\; \prime} )\right)
 \nonumber  \\[1.5ex]
&\;& \;\; \times (D_R J(p-Q) J^*(p-Q^\prime)) \nonumber \\[1.5ex]
&\approx& \int \frac{d^2 {\vQ_\perp}}{(2\pi)^2}\rho^{(0)}(p-Q)
\int \prod\limits_{i=1}^{n_s}  
\left( d^2 \vq_i \frac{d^2\sigma_{el}}{d^2\vq_{i\perp}} \right)
\left( \frac{d^2 {\vq_i}^{\; \prime}}{(2\pi)^2} T({\vq_i}^{\; \prime})
\right) \nonumber \\[1.5ex]
&\;& \;\; \times (2\pi)^2 
\delta^2(\vQ_\perp - \sum_{j=1}^{n_s} \vq_{i\perp})
\; , \qquad \label{rhoel0} 
\end{eqnarray} 
which includes the expected jet acoplanarity~\cite{Blaizot,Uli} 
resulting from random walk but no recoil energy loss.

The major simplification in the above derivation arises from the assumption
that $|\vq_i -\vq_i^{\;\prime}|\sim 1/R_\perp\ll \mu\ll \lambda$ so that
the  main source of non-diagonal dependence arises through the $T(\vq_{i\perp}
 -\vq_{i\perp}^{\;\prime})$ factors. This is what allows us to 
recover the above classical elastic parton cascade form in terms 
of the effective differential  cross sections
\begin{equation}
\frac{d^2\sigma_{el}}{d^2\vq_{i \perp}}= \frac{C_R C_2(i)}{D_A}
 \frac{4 \alpha_s^2}{(q_{i \perp}^2+\mu^2)^2} \; .
\label{sigel}\end{equation}
The elastic cross section is then
$\sigma_{el}= (C_R C_2(i)/D_A) (4\pi\alpha_s^2/\mu^2)$ which for a glue jet propagating
in a glue plasma with $\mu=0.5$ GeV is typically $\sigma_{el}\sim 2$ mb.
(See Ref.~\cite{ZPC} for a new Monte--Carlo implementation of such
 an elastic parton cascade.)

We will neglect  the influence of elastic multiple scattering
on the inclusive distribution of jets 
because $\rho^{(0)}(p)$ is broadly
distributed and the elastic energy loss is relatively small. 
In this case, $\rho^{(0)}(p-Q)\approx\rho^{(0)}(p)
$ can be factored out of eq.~(\ref{rhoel0}) leading to
\begin{eqnarray}
\rho^{(n_s)}_{el}(p)
&\approx&\rho^{(0)}(p) \;(\sigma_{el} T(0))^{n_s} 
\; , \qquad \label{rhoel} 
\end{eqnarray} 
which is simply proportional to the free space spectrum
times a geometrical factor 
(the  Glauber multiple collision probability).

\subsection{Gluon Radiation with Quantum Cascading}

The amplitude to emit a gluon together with the  jet including final state
interactions with $n_s$ scattering centers is given by
\begin{eqnarray}
 {\cal{M}}^{(n_s)}_{J}\otimes{\cal{M}}_{n_s}&=& \int d^4 P \; J(P) e^{i Px_0} 
\int \prod_{i=1}^{n_s} \left ( \frac{d^3 \vq_i}{(2\pi)^3} \;
(-2i P^0)  V(\vq_i) e^{-i \vq_i \vx_i} T(i)
\right )  \nonumber \\[1.5ex]
&\;& \times \;  {\cal M}_{n_s}(k,p;q_1,\cdots,q_{n_s}) 
\; \delta^4(P-k-p+\sum q_i) \;  .
\end{eqnarray}
This leads to the conditional double inclusive jet and gluon
probability distribution
$$d^{\, 6}{\cal D}^{(n_s)}_{rad}=\rho^{(n_s)}_{rad}(k,p) 
\frac{d^3\vk}{(2\pi)^32\omega}\frac{d^3\vp}{(2\pi)^3 2p_0} \; ,$$
\begin{equation}
\rho^{(n_s)}_{rad}(k,p)=\left \langle |{{\cal M}^{(n_s)}_J} 
\otimes{\cal{M}}_{n_s} |^2 
\right \rangle\equiv { 
(\prod_{i=1}^{n_s}\frac{1}{D_i})
\tr \left \langle |{{\cal M}^{(n_s)}_J} \otimes{\cal{M}}_{n_s} |^2 
\right \rangle_{\vec x} }  \; ,
\end{equation}
which involves averaging over all initial target colors and summing over
all final colors and the locations ${\vec x}_i$ of the target centers.
 
After integrating over the energy-conserving delta functions we arrive at
the momentum space distribution density
\begin{eqnarray}
\rho^{(n_s)}_{rad}(k,p)&=& \int \prod\limits_{i=1}^{n_s}  
\left (\frac{d^3 \vq_i}{(2\pi)^3} \, 
\frac{d^3 {\vq_i}^{\; \prime}}{(2\pi)^3} \; (2 p^0)^2 V(\vq_i)
V^*({\vq_i}^{\; \prime}) (C_2(i)/D_A) \right )  \nonumber \\[1.5ex]
&\;&  \times \;
{  \left \langle \begin{array}{c}   
{\displaystyle \prod\limits_{i=1}^{n_s} 
e^{-i (\vq_i-{\vq_i}^{\; \prime}) \vx_i} }   
\end{array} \right \rangle_{\vec x} }  
J(k+p-\sum q_i)J^*(k+p-\sum q_i^{\; \prime})
 \nonumber \\[1.5ex]
&\;& \times \; \tr\left({\cal M}_{n_s}(k,p;q_1,\cdots,q_{n_s}) 
{\cal M}_{n_s}^\dagger(k,p;q_1^{\; \prime},\cdots,q_{n_s}^{\; \prime}) 
\right) \; . \label{mm} 
\end{eqnarray}
The amplitude ${\cal M}_{n_S}$ is a sum over all time-ordered graphs involving
$n_s$ scatterings in which one gluon is radiated. 

For a  slowly varying source,  $J(p+k-Q)\approx J(p+k)$ can be factored
out of the  integral in eq.~(\ref{mm}) leading to
\begin{eqnarray}
\rho^{(n_s)}_{rad}(k,p)&\approx& |J(p+k)|^2 \int \prod\limits_{i=1}^{n_s}  
\left (\frac{d^3 \vq_i}{(2\pi)^3} \,
 \frac{d^3{\vq_i}^{\; \prime}}{(2\pi)^3} \; V(\vq_i)
V^*({\vq_i}^{\; \prime}) \frac{(2p^0)^2 C_2(i)}{D_A}
 T(\vq_{i \perp}-{\vq_{i \perp}}^{\; \prime} )\right )  
 \nonumber \\[1.5ex]
& \; & \times \; {  \left \langle \begin{array}{c}   
{\displaystyle \prod\limits_{i=1}^{n_s} 
e^{-i (q_{iz}-{q_{iz}^{\; \prime}) z_i }}}   
\end{array} \right \rangle_z } 
\nonumber \\[1.5ex]
& \; & \times  \; 
\tr\left({\cal M}_{n_s}(k,p;q_1,\cdots,q_{n_s}) 
{\cal M}_{n_s}^\dagger(k,p;q_1^{\; \prime},\cdots,q_{n_s}^{\; \prime}) 
\right) \; . \label{mm2} 
\end{eqnarray}
The integration over the $q_{iz}$ can again be performed by contour integration
leading to a sum over many terms with longitudinal phases set by energy 
differences as in ~\cite{MGXW,BDMPS}. In the eikonal limit, a simple shortcut 
to that calculation is possible when the scattering  of potentials localized
in the longitudinal direction at $z_i$ can be approximated by impulse 
interactions at times, $t_i\approx z_i/c$~\cite{BDMPS}. In that case, time 
ordered perturbation theory can be used to express eq.~(\ref{mm2}) as
\begin{eqnarray}
\rho^{(n_s)}_{rad}(k,p) & \approx &
|J(p+k)|^2 \int \prod_{i=1}^{n_s} \left ( \frac{d^2 \vq_i}{(2 \pi)^2} \,
\frac{d^2 \vq_i^{\; \prime}}{(2 \pi)^2} \; V(\vq_i)V^*(\vq_i^{\; \prime}) 
\frac{C_2(i)}{D_A}
T(\vq_{i \perp}-\vq_{i \perp }^{\; \prime} ) \right )
\nonumber \\[1.5ex]
&\times& \left \langle
\tr\left[\left(\sum_{m=0}^{n_s}\sum_{l=1}^{2^{m}}
 {\cal M}_{n_s,m,l}(k,p;\vq_{1 \perp },\cdots,
\vq_{n_s \perp})\right) \right. \right.\nonumber \\[1.5ex]
&\times&  \left. \left.
\left(\sum_{m^\prime=0}^{n_s}\sum_{l^\prime=1}^{2^{m^\prime}}
{\cal M}_{n_s,m^\prime,l^\prime}^{\dagger}
(k,p;\vq_{1 \perp }^{\; \prime},\cdots,
\vq_{n_s \perp}^{\; \prime})\right) \right]\right \rangle_t
\; ,\label{mm3}
\end{eqnarray}
where the sums go over the $2^{n_s+1}-1$
different time ordered diagrams 
that we list in the next sections. The reduced radiation amplitudes,
${\cal M}_{n_s,m,l}\;$, for a given $n_s$
are labeled by two integers, $0\le m\le n_s$ and $0\le l$.
Here $m$ specifies  the time interval $t_m<t<t_{m+1}$
in which the gluon is radiated and $l$ specifies
which of the $2^{n_s-m}$  final state interaction patterns 
 with centers $m+1,\cdots,n_s$ occurs. 

\subsection{Reduced Radiative Amplitudes}
 
It is convenient to associate with each amplitude diagram, 
${\cal M}_{n_s,m,l}$,
an $n_s$ dimensional binary array, $\vec{\sigma}$, that decodes
the label $l$ for a given $m$ as follows: 
\begin{equation}
\vec{\sigma}=(\sigma_1=0,\cdots,\sigma_m=0,
\sigma_{m+1},\cdots,\sigma_{n_s})_m \Longleftrightarrow
l= (\sum_{j=1}^{n_s}\sigma_j 2^{j})/2^{m+1}\;,
\end{equation}
where $ \sigma_i=0$ for $1\le i\le m$ while
$ \sigma_i=0$ or 1 for $m+1\le i\le n_s$
depending on the integer part of $l/2^{i-m-1}$.  
This is the base 2 representation of 
 $l\times 2^{m+1}=\sigma_{n_s}\cdots\sigma_1$ so that $0\le l\le 2^{n_s-m}-1$.
For a fixed $m$, $l=0$ corresponds then to no final state gluon rescattering,
while $l=2^{n_s-m}-1$ corresponds to the gluon rescattering
with all $n_s-m$
remaining scattering centers after $t_m$.

For example, Fig. 2 shows ${\cal M}_{5,1,10}$ as
the time ordered diagram with $n_s=5$ scattering centers, 
where gluon was emitted from the jet between $t_1$ and $t_2$
and $l=10$ corresponding to  $\vec{\sigma}=(0,0,1,0,1)$ encodes
the fact that the gluon received impulses $\vq_{\perp 3},\vq_{\perp 5}$
at times, $t_3$ and $t_5$.

The binary representation is useful as a mask to extract the
correct kinematic variables automatically. The transverse momentum 
of the gluon between interactions $i$ and $i+1$ is simply
\begin{equation}
\vK_i= \vk_\perp -\sum_{j=i+1}^{n_s} \sigma_j \vq_{j\perp}
=\vK_{i-1}+\sigma_i \vq_{i\perp}\;
\;, \label{kmask} \end{equation}
with $\vK_{n_s}\equiv \vk_\perp$, the final gluon transverse momentum.
For the example in Fig. 2, $\vK_0=\vK_1=\vK_2=\vk_\perp-\vq_{3\perp}-
\vq_{5\perp}$ and
 $\vK_3=\vK_4=\vk_\perp-\vq_{5\perp}$.

\begin{center}
\vspace*{7.0cm}
\includegraphics{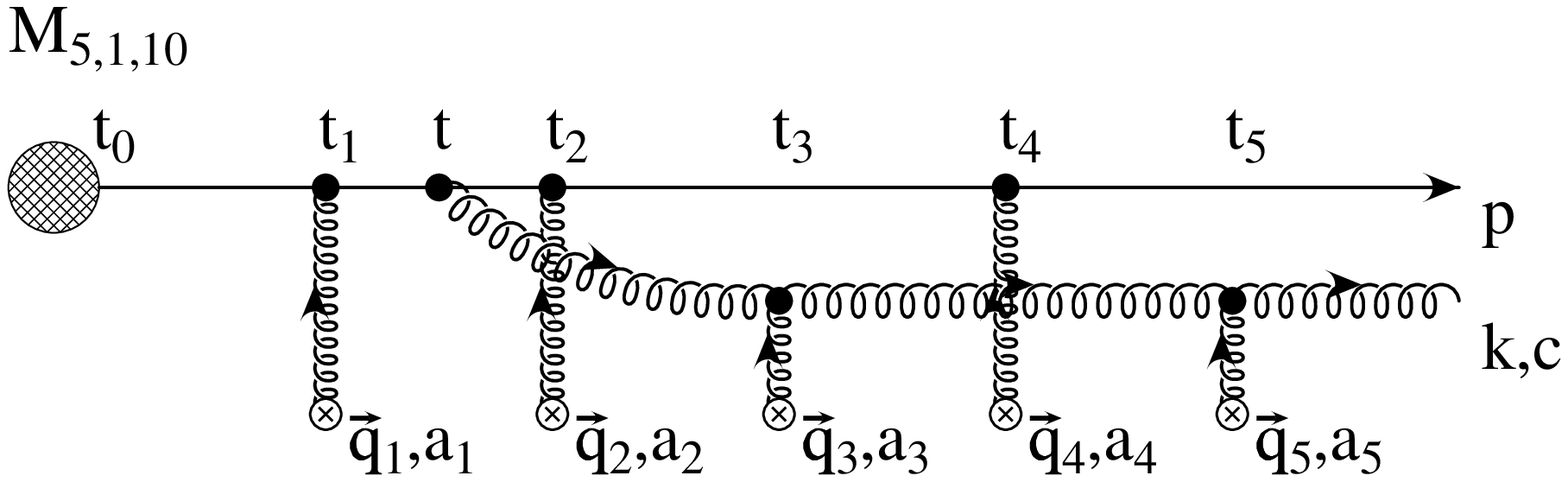}
\vskip -55pt
\begin{minipage}[t]{15.0cm}
{\small {\bf Fig.~2.}
Example of  the reduced {\cal M}$_{5,1,10}$ radiative amplitude
for $n_s=5$.}
\end{minipage}
\end{center}
\vskip 4truemm

The complete amplitude matrix can be constructed in a few steps.
First, the time integration over the emission times between $t_m$ 
and $t_{m+1}$ and the $n_s$ impulses gives rise to the factor
\begin{equation}
\frac{\omega}{\omega_{ml}}
\left(e^{i\omega_{ml} t_{m+1}}-e^{i\omega_{ml} t_{m}}\right)
 e^{i\Phi_{lm}} \; \;, \label{phases} 
\end{equation}
where the energy shift at the emission point is denoted 
\begin{equation} 
\omega_{ml}= \frac{K_m^{\,2}}{2\omega}=\frac{\left(\vk_\perp - \sum_{j=1}^{n_s}
\sigma_j\vq_{j\perp}\right)^2}{2\omega}
\;. \label{energy} 
\end{equation}
The apparent singularity at $\omega_{ml}=0=K_m^{\,2}$ 
due to the energy denominator is cancelled
for finite $t_{m+1}-t_m <\infty$. The singular $t_{n_s+1}=\infty$
phase is  always suppressed by an adiabatic damping factor (not shown).
The only singular amplitude is in fact, the one corresponding
to the factorization limit in which final state interactions factor out.
Including the eikonal gluon emission vertex $2 i g_s \epsilon_\mu(K_m)p^\mu
\approx g_s \vec{\epsilon}_\perp\cdot \vK_m$,
the special case for $m=n_s$, $l=0$  reduces to
$${\cal M}_{n_s,n_s,0}= - 2 i g_s {\vec{\epsilon}_\perp \cdot
\vK_{n_s} \over K^2_{n_s} } e^{i t_{n_s} {K^2_{n_s} \over 2\omega}}  
c (a_{n_s}\cdots a_1) = e^{i (t_{n_s}-t_0) {k^2_\perp 
\over 2\omega}} {\cal M}_0 \;(a_{n_s}\cdots a_1)\; \; ,$$
where ${\cal M}_0$ is given by eq.~(\ref{m01}).
All the other amplitudes remain bounded 
$|{\cal M}_{n_s,m,l}|< C (t_{m+1}-t_m)$.
On the other hand, in the formal $t_0\rightarrow -\infty$ limit, 
all the $2^{n_s}-1$ singularities at $\omega_{ml}=0$ 
become exposed. 

The overall eikonal phase, $\Phi_{ml}$, acquired by the gluon through
final state interactions is
\begin{eqnarray}
\Phi_{ml}&=&\sum_{j=1}^{n_s}\Omega_{mlj} t_j 
= \sum_{j=1}^{n_s}\frac{t_j}{2\omega}\left(\vK_j^{\,2}-\vK_{j-1}^{\,2}\right)
= \sum_{j=m+1}^{n_s} \sigma_j \frac{t_j}{2\omega}\left(\vK_j^{\,2}-
(\vK_{j}-\vq_{j\perp})^2\right) \; , \qquad \label{phiml} 
\end{eqnarray}
where $\Omega_{mlj}=\partial \Phi_{ml}/\partial t_j$.

The amplitude for color exchange leaving the final gluon
with color $c=1,\cdots, D_A$  is specified by a particular 
color matrix, $T_{mlc}$.  The scattering of the jet with center $i$
brings in a color matrix $a_i\equiv T_{a_i}(R)$ in the $R$ 
representation. Gluons interact with center $i$,
on the other hand with $T_{a_i}(A)$ in the adjoint
representation. It is numerically convenient, however,
to express the complete color structure involving gluon 
final state interactions in terms
of commutators in  the jet ($R$ dim) representation.
Label sequentially in increasing order all $n_{g}=\sum_k \sigma_k$
color matrices, $a_k$ where $\sigma_k=1$, by $c_1,\cdots, c_{n_{g}}$. 
Label  all remaining $n_s-m-n_{g}$ matrices with $k>m$ and 
$\sigma_k=0$ by $b_1,\cdots,b_{n_s-m-n_{g}}$. Then
\begin{eqnarray}
T_{mlc}=\left(b_{n_s-m-n_{g}}\cdots b_1\right)
\left[ \cdots [c,c_{n_{g}}],\cdots,c_1\right]\left(a_m\cdots a_1
\right)
\;. \label{tmlc} \end{eqnarray}
In the example of Fig. 2, $T_{1,10,c}=a_4 a_2[[c,a_5],a_3] a_1$.
Eq.~(\ref{tmlc})  defines a straightforward algorithm
for  numerical computation of color factors, 
$$C_{mlm^\prime l^\prime}=\sum_{c,a_1,\cdots, a_n}
\tr \;( \, T_{mlc} T^\dagger_{m^\prime l^\prime c}\, )\;,$$ 
needed as weights of the different interference terms in eq.~(\ref{mm3}) .

Finally, the reduced radiation amplitudes with final state interactions
are given by
\begin{eqnarray}
{\cal M}_{n_s,m,l}=2ig_s\frac{\vec{\epsilon}_\perp \cdot \vK_m}{
K_m^2}\left(e^{i\frac{t_{m+1}K_m^2}{2\omega}}-e^{i
\frac{t_{m}K_m^2}{2\omega}}\right)
 e^{i\Phi_{lm}} T_{mlc} \; . \label{exact}
\end{eqnarray}

\subsection{Interference Form Factors}

If the ``width'', $R_\perp\gg \lambda\gg 1/\mu$, of the target
transverse  to the jet direction is large compared to its thickness,
$L$, (i.e., $R_\perp\gg L> \lambda\gg 1/\mu$),  then the off diagonal
dependence in $|\vq_{i\perp}-\vq_{i\perp}^{\;\prime}|\sim 1/R_\perp $ 
in eq.~(\ref{mm3}) can be ignored in all factors except 
$T(\vq_{i\perp}-\vq_{i\perp}^{\;\prime})$
(see  Sec.~4.1 for more detailed discussion on this).
In this case, the interference terms arising from the distribution of
scattering times, $t_i$,  involve the following form factors:
\begin{eqnarray}
F_{mlm^\prime l^\prime}\equiv \left\langle \;
e^{i(\Phi_{ml}-\Phi_{m^\prime l^\prime})}\left(e^{i t_{m+1}\omega_{ml}}
-e^{i t_{m}\omega_{ml}}\right)
\left(e^{-i t_{m^\prime+1}\omega_{m^\prime
l^\prime}}-
e^{-i t_{m^\prime} \omega_{m^\prime l^\prime}}\right)\;\right\rangle
\; .\label{formf}
\end{eqnarray}
For $m,m^\prime<n_s$, the form factors insure
that $F_{mlm^\prime l^\prime}/(\omega_{ml}\omega_{m^\prime
l^\prime})$ have no singularities except for $m,m^\prime=n_s$,
and control the magnitude of the destructive LPM interference
effects. The diagonal form factors are given by
\begin{eqnarray}
F_{mlml}&=& 2-2\,\langle \; \cos((t_{m+1}-t_m)\omega_{ml})
\;\rangle \; ,\label{diag }
\end{eqnarray}
except that  $F_{n_s0n_s0}=1$  since $t_{n_s+1}=\infty+i\epsilon$.
The extreme  limits of quantum cascading
correspond to (1) the incoherent Classical Parton Cascade
limit and (2) the  coherent Factorization
limit. In the first case, all non-diagonal form factors
vanish, while in the second case all but the $m=n_s$
diagonal form factor vanish:
\begin{eqnarray}
F_{mlm^\prime l^\prime}&=& 2 \delta_{m,m^\prime} \delta_{l,l^\prime} 
\;\; {\rm Classical\; Parton \;Cascade}\; , \nonumber \\
F_{mlm^\prime l^\prime}&=& \delta_{m,n_s} \delta_{m^\prime,n_s}
\;\; {\rm Coherent\; Factorization}
\; .\label{limformf}
\end{eqnarray}

Simple analytic expressions can be obtained for the form factors
if the ensemble average over the interaction times
is taken as in eq.~(\ref{zdis}).  The form factors involve a sum of four 
terms of the form
\begin{eqnarray}
 \left\langle\; e^{i\sum_{j=1}^{n_s} 
(t_j - t_{0})\omega_j} \; \right \rangle&=&
\left\langle \; e^{i\sum_{j=1}^{n_s} 
\tau_j \tilde \omega_j } \; \right \rangle
=\prod_{j=1}^{n_s} \int_{0}^\infty \frac{ d\tau_j}{\lambda_j} 
e^{-\frac{\tau_j}{\lambda_j} (1-i\lambda_j\tilde\omega_j)} \;
\nonumber \\[1.5ex]
&=&\prod_{j=1}^{n_s} \frac{1}{1-i\lambda_j
(\omega_j+\cdots+\omega_{n_s})}\; ,\label{tmodel}\label{expff}
\end{eqnarray} 
where $\tau_j\equiv t_j-t_{j-1}$, $\tilde\omega_j=\omega_j+\cdots
+\omega_{n_s}$. Here $\lambda_j=\langle \tau_j\rangle$ 
is the average distance between 
the $j-1$ and $j^{th}$ scattering centers.
This ensemble is characterized by the following distribution 
of $t=t_k-t_{0}$:
\begin{eqnarray}
P_k(t)&=& \langle \;\delta(t_k-t_0-t) \;\rangle=
\int \frac{d\omega}{2\pi} e^{-i\omega(t_k-t_0)} \prod_{j=1}^k
\frac{1}{1-i\omega\lambda_j} 
\nonumber \\[1.5ex]
&=& 
\sum_{j=1}^{n_s} \frac{e^{-t/\lambda_j}}{\lambda_j} 
\left(\prod_{l\ne j}^{k}\frac{1}
{(1-\lambda_l/\lambda_j)}\right)\;
\quad \stackrel{\lambda_k=\lambda}{\longrightarrow}
\quad \frac{e^{-t/\lambda}}{\lambda}
\frac{(t/\lambda)^{k-1}}{(k-1)!}\label{pit}
\; .
\end{eqnarray} 
The geometry of the plasma is particularly simple if
$\lambda_k=\lambda$, with 
$$\langle \, (t_k-t_0)\,\rangle =k\lambda \;, \qquad
\langle \, (t_j-t_0)(t_k-t_0)\, \rangle =k (j+1)\lambda^2\;, \;\; j\ge k\;.$$

For   more realistic ensembles corresponding
to an expanding target with centers distributed according to $\rho(z,t)$.
The form factors must be evaluated numerically via
\begin{eqnarray}
\left\langle\; e^{i\sum_{j=1}^{n_s} (t_j-t_0) \omega_j}\;\right \rangle
&=& \frac{1}{P_n}
\int_{t_0}^\infty dt_n e^{-\chi(\infty,t_n)}
\int_{t_0}^{t_n} dt_{n-1}
\cdots\int_{t_0}^{t_2} dt_1 \nonumber \\[1.5ex]
&&\times \left(\;\prod_k \,\sigma_{el}\, 
\rho(z_k=t_k, t_k)\, e^{i(t_j-t_0) \omega_j}\,e^{-\chi(t_k,t_{k-1})}
\; \right) \; , \nonumber \\[2.ex]
\chi(t,t^\prime)&=&\int_{t^\prime}^t d\;{\hat t}\; \sigma_{el} 
\rho(z={\hat t}, {\hat t}\,)
\;,\nonumber\\
P_{n_s}&=& e^{-\chi(\infty,t_0)}\frac{\chi^{n_s}(\infty,t_0)}{n_s!}
\; .\label{chi}
\end{eqnarray} 
Here  $\chi(t,t^\prime)$ is the average number
of interactions a jet moving with the speed of light
through the expanding target suffers between $t^\prime$ and $t$
with  $\sigma_{el}$  given by eq.~(\ref{sigel}).  Note that  $P_{n_s}$ is the
probability that only $n_s$ out of a large possible $N\gg 1$ target 
interactions occur in the medium.
In this more general case, the interactions are distributed as 
\begin{eqnarray} 
P_k(t_k) &=&  P_{n_s} \frac{d\gamma(t_k)}{dt_k} 
(1-\gamma(t_k))^{n_s-k} \gamma^{k-1}(t_k)
\; ,\label{piti} \nonumber \\[1.5ex]
\gamma(t)&=& \chi(t,t_0)/\chi(\infty,t_0)
\; .\label{gamt}
\end{eqnarray} 

For a static uniform slab of thickness $L$ with density $\rho_0 \theta(L-z)$ 
\begin{eqnarray}
\left\langle\; e^{i\sum_{j=1}^{n_s} t_j \omega_j}\;\right \rangle
&=& \frac{n_s!}{L^{n_s}}
\int_{0}^L dt_{n_s} e^{it_{n_s}\omega_{n_s}}
\cdots\int_{0}^{t_2} dt_1  e^{it_1\omega_1}
\; ,\label{slab}
\end{eqnarray} 
It is instructive to compare eq.~(\ref{slab}) with
eq.~(\ref{tmodel}). In the slab geometry setting $t_0=0$
$$\langle \, t_k \, \rangle=\frac{ k\; L}{(n_s+1)} \;, \qquad 
\langle \, t_j t_k \, \rangle=
\frac{k(j+1)\;L^2}{(n_s+2)(n_s+1)}\;, \;\; j \ge k \;.$$ 
Thus, if we set
$\lambda\equiv L/(n_s+1)$ 
$$\langle \, t_k \, \rangle=k\lambda \;, \qquad 
\langle \, t_j t_k \, \rangle=
k(j+1)\lambda^2 \frac{(n_s+1)}{(n_s+2)}\; , \;\; j \ge k \;.$$ 
For $n_s\gg 1$ the geometry of both ensembles
is quite similar if set $\lambda$ appropriately. 
Near the factorization limit, where $\sum_{j=1}^{n_s} \lambda \omega_j \ll 1$
we can expand
\begin{eqnarray}
&\;&  \left\langle\; e^{i\sum_{j=1}^{n_s} (t_j - t_{0})\omega_j}\;\right \rangle
\approx 1+ i\lambda\sum_{j=1}^{n_s} j\omega_j - r\;\lambda^2
\sum_{j=i}^{n_s} \sum_{k=1}^j k(j+1) \omega_j \omega_k 
\; ,\label{nearfac}
\end{eqnarray} 
where the only difference between eq.~(\ref{tmodel}) and eq.~(\ref{slab})
is that $r=1$ in the first case and $r=(n+1)/(n+2)$ in the second case.
The deviations from factorization can therefore be reasonably
well approximated by our simpler ensemble.
On the other hand, for large $\lambda\omega_j$, 
the slab form factors oscillate due to the sharp boundary
while the form factors in the
eq.~(\ref{tmodel}) ensemble decrease monotonically.
The deviations from  the classical parton cascade limit is
therefore more sensitive to the detailed form of the geometry of the medium.
We will continue to use below eq.~(\ref{tmodel}) to 
reduce the complexity of the calculation.

Since the final pattern of jet quenching  depends on the actual
distributions and correlations among scattering centers
in general, one should evaluate eq.~(\ref{formf}) numerically
for any actual application to nucleus-nucleus collisions~\cite{GLVII}.

\subsection{Relative Conditional Probabilities}

It is tempting to identify as in eqs.~(\ref{rhog0},\ref{dn0})
\begin{eqnarray}
R_{g}^{(n_s)}(\vk) = \frac{1}{16\pi^3}\frac{\rho^{(n_s)}_{rad}(\vk,\vp)}
{\rho^{(n_s)}_{el}(\vp+\vk)}  \;  
\end{eqnarray} 
as the invariant soft gluon number distribution associated with $n_s$
interactions.  However, because the radiated 
gluon can rescatter in addition to the incident jet, the conditional double
inclusive density, $\rho^{(n_s)}_{rad}$, cannot be factored into a product of
the elastic distribution and the gluon number distribution as in the case of
QED.  The problem is that the production and gluon rescattering amplitudes
interfere, and $\rho^{(n_s)}_{rad}$ includes  the process where the
gluon is radiated before the first collision and the $n_s$ recoil target
partons are produced by elastic scattering with the produced gluon rather than
the incident jet.  Therefore, we interpret $R^{(n_s)}_g$ as a relative
conditional probability density to find a gluon in the final state with
momentum $k$ given a final observed jet with momentum $p$ and given
that $n_s$ target
recoil partons are observed (with any momentum).

Let $P_{n_s}$ be the {\em a priori} probability that there are
$n_s$ scattering centers or target recoil partons.
The final jet quenched spectrum is then given by
\begin{eqnarray}
\rho_J(\vp\,)=\sum_{n_s} \frac{P_{n_s}}{Z_{n_s}}\left(
\rho^{(0)}(\vp\,) +  \int 
\frac{d^3 \vk}{\omega} \; \rho^{(0)}(\vp+\vk) 
\; R^{(n_s)}_g(\vk\,) \right)
\; ,\label{rhoj} 
\end{eqnarray}
where $\rho^{(0)}(p)$ is the initial hard jet distribution in 
eq.~(\ref{jet0}), and where the (wavefunction) renormalization factor
\begin{eqnarray}
Z_{n_s}=1+\int
\frac{d^3 \vk}{\omega}  R^{(n_s)}_g(\vk\,) 
\label{z}
\end{eqnarray}
insures that the integrated total number of jets is conserved.
It in fact corresponds to the first order in $\alpha_s$
term in the expansion of the inverse Sudakov factor
for the probability that no radiation accompanies the jet.
In this paper we only compute, $R^{(n_s)}_g(\vk\,)$, which
limits the discussion to single gluon emission: 
$R^{(n_s)}_g(\vk\,)\equiv R^{(n_s,n_g=1)}_g(\vk\,)$.

Unfortunately, even in the $n_s=0$ case the single
gluon approximation is not accurate because 
\begin{eqnarray}
Z_0= 1+ \frac{C_R\alpha_s}{\pi} \log^2(E_0/\mu) 
\; ,\label{z0}   
\end{eqnarray}
can be significantly bigger than one, indicating that multigluon
emission from the hard vertex is necessary. Summing the double logarithms
leads to a modified Bessel function 
multiplicity distributions as discussed in \cite{Field}.
The DLA leads to  \cite{Field},
\begin{eqnarray}
Z_0 &=& 1/I_0(r(E))\;, \nonumber\\[.5ex]
r(E)&=&\sqrt{8 C_R \alpha/\pi} \log({E/\mu})\;,\nonumber \\[.5ex]
\langle n_g\rangle&=& \frac{r}{2}\frac{I_1(r)}{I_0(r)}\approx \frac{r(E)}{2}
\;\; .
\label{multg}
\end{eqnarray}

In the general case of multiple gluon emission, as  required
in practical applications, we must  generalize eqs.~(\ref{rhoj},\ref{z}) as 
\begin{eqnarray}
\rho_J(\vp\,)=\sum_{n_s}  \frac{P_{n_s}}{Z_{n_s}}\left(
\rho^{(0)}(\vp\,) +  \sum_{n_g=1}^\infty\int 
\prod_{i=1}^{n_g} \frac{d^3 \vk_i}{\omega_i} \; \rho^{(0)}(\vp
+\sum_j \vk_j\,) 
 R^{(n_s,n_g)}_g(\vk_1,\cdots,\vk_{n_g}\,)  \right)
\; ,\label{rhojf} 
\end{eqnarray}
where the (wavefunction) renormalization factor becomes
\begin{eqnarray}
Z_{n_s}=1+ 
\sum_{n_g=1}^{\infty} \int 
\prod_{i=1}^{n_g} \frac{d^3 \vk_i}{\omega_i}
 \;  R^{(n_s,n_g)}_g(\vk_1,\cdots,\vk_{n_g}) 
\; .\label{zf}\end{eqnarray}
The implementation of this approach requires Monte--Carlo
methods that will be discussed in the subsequent paper \cite{GLVII}.

{}From the above formulas, we can infer that
the exclusive $n_g$ gluon number distributions associated
with $n_s$ scattering is 
\begin{eqnarray}
\omega_1\cdots \omega_{n_g}
\frac{dN^{(n_s,n_g)}_g}{d^3 \vk_1\cdots d^3 \vk_{n_g}} =
\frac{1}{Z_{n_s}} 
 R^{(n_s,n_g)}_g(\vk_1,\cdots,\vk_{n_g}) 
\; .\label{dng} 
\end{eqnarray}

\section{Case of One Scattering Center}

We consider in this section the simplest case
of gluon radiation in hard processes followed by one rescattering.

\subsection{Reduced Amplitudes and Transverse Profile Dependence}

The three graphs that generalize
the Gunion-Bertsch analysis~\cite{GUNION} to this  case
are shown in Fig. 3.

\begin{center}
\vspace*{8.8cm}
\includegraphics{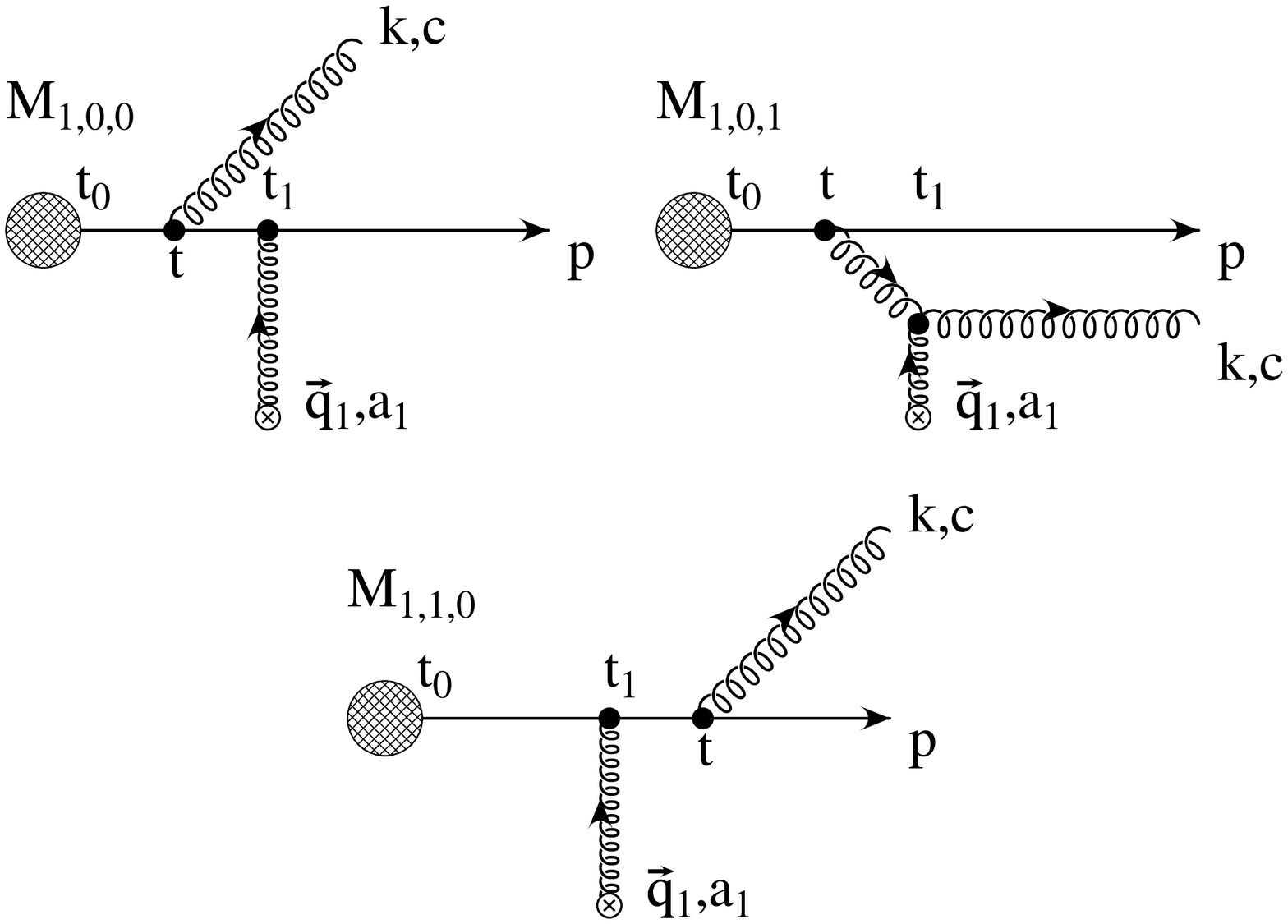}  
\vskip -20pt
\begin{minipage}[t]{15.0cm}
{\small {\bf Fig.~3.}
Three contributions to the soft gluon radiation amplitude  {\cal M}$_J \otimes$
{\cal M}$_1$ in case of one scattering.}
\end{minipage}
\end{center}
\vskip 4truemm

Following BDMPS~\cite{BDMPS}, the three radiative 
amplitudes are given by
\begin{eqnarray}
{\cal{M}}_{1,0,0} &=& 2 i g_s {\vec{\epsilon}_\perp \cdot
\vk_\perp \over k^2_\perp } 
(e^{i t_1 {k^2_\perp \over 2\omega}} 
-e^{i t_0 {k^2_\perp \over 2\omega}} ) a_1 c \; , 
\label{m001}\\[.5ex]
{\cal{M}}_{1,0,1} &=& 2 i g_s {\vec{\epsilon}_\perp \cdot
(\vk - \vq_1 )_\perp \over (k - q_1)^2_\perp } 
e^{i t_1 {k^2_\perp - (k - q_1)^2_\perp \over 2\omega}} 
\times \nonumber \\[.5ex]
 \qquad && \times (e^{i t_1 {(k - q_1)^2_\perp \over 2\omega}} 
-e^{i t_0 {(k - q_1)^2_\perp \over 2\omega}} ) 
 \left [ c, a_1 \right ] \; , \\[.5ex]
{\cal{M}}_{1,1,0} &=& 2 i g_s {\vec{\epsilon}_\perp \cdot
\vk_\perp \over k^2_\perp } 
( -e^{i t_1 {k^2_\perp \over 2\omega}} ) c  a_1\; . 
\label{m1}
\end{eqnarray}
Here $t_1$ is the impulse time when $\vq_{1 \perp}$ transverse
momentum is imparted to the jet relative to its axis $\vp$. The color
charge flow is modified by an additional matrix $a_1$, that appears in
different orderings relative to the radiation vertex $c$ in these
amplitudes. A new infrared singularity $\vk_\perp - \vq_{1 \perp}=0$
emerges due to the dipole radiation associated with the exchanged
gluon. However, the phase factors cancel at that point and the amplitude 
remains finite proportional to $t_1-t_0$.

To keep the somewhat cumbersome notation compact we introduce the
following reduced amplitudes
$$\vec{H}={\vk_\perp \over k^2_\perp }\; , \qquad
\vec{C}_{(i_1i_2 \cdots i_m)}={(\vk - \vq_{i_1} - \vq_{i_2}- 
\cdots -\vq_{i_m} )_\perp 
\over (k - q_{i_1} - q_{i_2} - \cdots -q_{i_m} )^2_\perp } \;, $$
$$\vec{B}_i \equiv \vec{H} - \vec{C}_i \; , \qquad
\vec{B}_{(i_1 i_2 \cdots i_m )(j_1j_2 \cdots i_n)} \equiv 
\vec{C}_{(i_1 i_2 \cdots j_m)} - \vec{C}_{(j_1 j_2 \cdots j_n)}\; ,$$
\vskip 5pt
\noindent which we call ``hard'', ``cascade'', ``Gunion--Bertsch'', and 
``Gunion--Bertsch cascade''. The time integrals involve
energy differences
$$\omega_0 = k_\perp^2/2\omega=1/t_f\;  , \quad 
\omega_{(i_1 i_2 \cdots i_m )} = (k- \vq_{i_1} - \vq_{i_2}- 
\cdots -\vq_{i_m} )_\perp^2/2\omega \; ,  $$
$$ \omega_{0i}\equiv \omega_0-\omega_i\; , \quad 
\omega_{(i_1 i_2 \cdots i_m )(j_1j_2 \cdots i_n)} \equiv
\omega_{(i_1 i_2 \cdots i_m )}-\omega_{(j_1j_2 \cdots i_n)} \;,  $$
\vskip 5pt
\noindent that control the formation physics. A quantity with two indices is
defined to be the difference of two quantities with a single index
(note that an array in parentheses counts as a single index). 
Thus, for example $t_{10}\equiv t_1- t_0,\; \omega_{10}\equiv
\omega_{1} - \omega_{0}$.  

To make contact with the Gunion-Bertsch bremsstrahlung radiation 
distribution~\cite{GUNION} and facilitate the physics interpretation 
we rearrange the sum of  the three amplitudes as follows:
\begin{equation}
{\cal{M}}_1 = -2 i g_s  e^{i t_0 \omega_0} \vec{\epsilon}_\perp  
\cdot \left\{ \vec{H} a_1c + \vec{B}_1  e^{i t_{10} \omega_0}
\left [ c, a_1 \right ] + \vec{C}_1  e^{-i t_{10}( {\omega_1-\omega_0)})}
\left [ c, a_1 \right ] \right\} .
\label{m1f}
\end{equation}  
Including an adiabatic damping factor $e^{-\epsilon t_1}$, 
we see that for $t_{1}\rightarrow \infty$
the pure hard radiation formula for the $n_s=0$ case is recovered.
\begin{equation}
\lim_{t_1\rightarrow+\infty}{\cal{M}}_{1} =  a_1\;{\cal{M}}_0 \; .
\end{equation}
On the other hand, in the $t_0\rightarrow -\infty$ limit,
\begin{equation}
{\cal{M}}_{1} \rightarrow -2 i g_s \vec{\epsilon}_\perp 
\cdot \vec{B}_1  e^{i t_{1} \omega_0}
\left [ c, a_1 \right ] = {\cal{M}}_{GB} \; ,
\end{equation}
corresponding to the isolated bremsstrahlung amplitude. This leads to the 
characteristic GB radiation spectrum
\begin{equation}
\frac{dN^{(GB)}_g}{dyd^2\vk_\perp}= C_A \frac{\alpha_s}{\pi^2}\frac{q_1^2}
{k_\perp^2(k -q_1)_\perp^2} \; ,
\label{BGrad}\end{equation}
that has {\bf to be averaged} over the transfered momentum. Here only the 
adjoint Casimir $C_A=N_c$ enters and the result is independent  of $C_R$. Note 
that in this limit, the $\vk_\perp =\vq_{1 \perp} $ is exposed  and 
in~\cite{GUNION} is regulated with an appropriate hadronic form factor. 

For the theoretical application to a parton penetrating through infinite 
matter, this singularity is regulated by the in-medium self energy of the 
gluon, $\Pi^{\mu\nu}(k)$. The transverse plasmon modes satisfy $\omega^2=
k^2+\Pi_T(\omega,k)$ in an infinite plasma medium, with $\Pi_T(\omega_{pl}(k)
,k)\approx \omega_{pl}^2\approx \mu^2/3$ and $\mu^2=g^2 T^2(2N_c+N_f)/6$ is 
the Debye screening mass in thermal pQCD. For $k_z\gg T$ on the other hand, 
$\omega_{pl}(k)\approx k_z + (k_\perp^2+\mu^2/2)/(2 k_z)$. Thus the infrared 
singularities are regulated by a scale $\sim \mu^2$. Near the light cone, the 
one loop approximation for the self energy is not likely to remain accurate 
though. In fact, gauge invariance requires that the same $\mu^2$ screening 
scale regulate both the potential singularity and the above infrared 
divergences~\cite{MGXW,BDMPS}.

In contrast to the above, $t_0\rightarrow -\infty$ limit, the 
$\vk_\perp =\vq_{1 \perp}$ singularity is also regulated by canceling 
phases for $t_{10}$ finite. In this case of interest here, it is the maximum
of $\mu^2$ and $\omega/t_{10}$ that regulates that singularity. For large 
$\omega\gg \omega_{BH}=\mu^2 \lambda/2$, most important for energy loss, the 
latter scale in fact sets the characteristic transverse momentum of the
gluon in the $k_\perp >\mu $ domain.

Summing over final polarizations and colors and averaging over initial colors, 
the $n_s=1$ gluon distribution from eqs.~(\ref{mm3},\ref{m001}-\ref{m1}) is
\begin{eqnarray}
\rho^{(1)}_{rad}(k,p)&\approx &  |J(p+k)|^2 \int \frac{d^2 \vq_1}{(2\pi)^2} \, 
\frac{d^2 \vq_1^{\; \prime}}{(2\pi)^2}  V(\vq_1) V^*(\vq_1^{\; \prime}) 
\frac{C_2(i)}{D_A}T(\vq_{1 \perp}-\vq_{1 \perp }^{\; \prime} )  
\nonumber \\[1.ex]
&&\times \;  {\displaystyle \left \langle \begin{array}{c}
 \tr\, \left(  \sum\limits_{1,m,l,m^\prime,l^\prime} {\cal M}_{1,m,l}
(k,p;\vq_{1 \perp}) {\cal M}_{1,m^\prime,l^\prime}^\dagger 
(k,p;\vq_{1 \perp}^{\; \prime}  )
\right )  \end{array}  \right \rangle_t } \; .\qquad \label{rho1} 
\end{eqnarray}

The main complexity in the above expression arises from the mixing of
different $\vq_1$ and $\vq_1^{\; \prime}$ components. However, if the
transverse profile $T(\vx_\perp)$ of the target is approximately a
constant over dimensions $R\gg 1/\mu$, then the difference
$|\vq_1-\vq_1^{\;\prime}|\sim 1/R$ can be neglected in {\em some} of
the terms. The only sensitive dependence of that difference arises
through the phase factors in eq.~(\ref{m001}-\ref{m1}). A systematic 
expansion in that difference is possible changing variables to
$\vec{Q}=(\vq_1+\vq_1^{\;\prime})/2$ and $\vq=\vq_1-\vq_1^{\;\prime}$
and expanding in $\vq$. To leading order in this expansion
\begin{eqnarray}
\rho^{(1)}_{rad}(k,p)&=&  4 g_s^2 \, |J(p+k)|^2 \int \frac{d^2 \vQ}{(2\pi)^2} 
\frac{C_2(i)}{D_A}|V(\vQ)|^2
\int \frac{d^2\vq }{(2\pi)^2}\; T(\vq\,) \times   \nonumber \\[.5ex]
& &  {\displaystyle  \; \begin{array}{l} \left \langle 
\tr\; \left(  (\vec{H} a_1c + \vec{B}_Q  e^{i t_{10} \omega_0}
\left [ c, a_1 \right ] )\cdot (\vec{H} c a_1 + \vec{B}_Q  
e^{-i t_{10} \omega_0} \left [ a_1,c \right ] )\right) 
\right.  \\[2ex]
+ \tr\, \left( |\vec{C}_Q|^2  e^{-i t_{10}(\omega_1-\omega_1^\prime)}
\left [ c, a_1 \right ] \left [ a_1, c \right ] \right)
 \\[2ex]
 + \tr\, \left(  (\vec{H} a_1c + \vec{B}_Q  e^{i t_{10} \omega_0}
\left [ c, a_1 \right ] )\cdot \vec{C}_Q  e^{+i t_{10}
( \omega_1-\omega_0)} \left [ a_1,c \right ] \right) 
 \\[2ex]
\left.  + \tr\, \left(  (\vec{H} a_1c + \vec{B}_Q  e^{-i t_{10} \omega_0}
\left [ c, a_1 \right ] )\cdot \vec{C}_Q  e^{-i t_{10}
( \omega_1^\prime-\omega_0)} \left [ a_1,c \right ] \right ) 
\, \right \rangle_{t_1} \; \; . \end{array}  } 
\label{rho12} 
\end{eqnarray}

Note that $\omega_1=(\vk-\vQ-\vq/2)^2/2\omega,\; 
\omega_1^\prime=(\vk-\vQ+\vq/2)^2/2\omega$, and therefore, 
the extra phase factor multiplying the diagonal 
cascade contribution has a phase
$$ \vec{b}_1\cdot\vq=t_{10}(\vk-\vQ)\cdot\vq/\omega \; ,$$ where we
drop ${\cal O}(t_{10}\mu^2/\omega)$ terms. The integration over
$d^2\vq$ transforms $T(\vq \,)$ therefore into transverse spatial
profile $T(\vec{b}_1)$ evaluated at the impact parameter
$$\vec{b}_1=t_{10}(\vk-\vQ)/\omega \; ,$$ at which a gluon
produced at the origin with transverse momentum $\vk-\vQ$ would pass
by the scattering center at $z_1=c t_1$. This tends to suppress the
cascade contribution as intuitively expected for kinematic
configurations where the gluon ``misses'' the target. On the other
hand, the diagonal hard and Gunion-Bertsch contributions remain
unsuppressed since the rescattering of the jet occurs at zero impact
parameter in the high energy limit. For a broad target, the
 transverse profile is approximately a constant
$T(\vb_1)\approx T(0)\sim 1/\pi R^2$.
However, in general the cascade term is modulated by 
 a relative target transverse profile  factor
 $${\cal T}(\vb_1)\equiv T(\vb_1)/T(0) \; .$$ 
The relative conditional probability
for gluon radiation  in this $n_s=1$ case for a 
fixed momentum transfer $\vQ_\perp \rightarrow \vq_{1 \perp }$
therefore reduces to
\begin{eqnarray}
R^{(1)}_g(\vk,\vq_{1\perp})  &=& C_R {\alpha_s \over \pi^2} 
\left\{\vec{H}^2 + R\,(\vec{B}^2_1 +{\cal T}(\vb_1)\; \vec{C}^2_1  )
- R\left ( \vec{H} \cdot \vec{B}_1 \cos (t_{10}\omega_0) 
\right) \right.  \nonumber \\[.5ex] 
&& -  R\, {\cal T}(\vb_1/2 ) \left. \left(\vec{H}\cdot\vec{C}_1 
\cos (t_{10} \omega_{10}) - 2\vec{C}_1\cdot\vec{B}_1
\cos (t_{10}\omega_1) \right) \right\} \,  \; . \label{dn10}
\end{eqnarray}
Note that the transverse profile factor modulates
 the interference term
between ${\vec C}$ and ${\vec H}+{\vec B}$ at the average impact
parameter, $\vb_1/2$ of the jet and gluon at the scattering
center. Note that for $t_{10} \rightarrow 0$ there is a perfect
cancellation of all but the hard radiation term
since $b_1 \propto t_{10} \rightarrow 0$ and $
{\cal T}(b=0)=1\;$. In fact, this factorization limit
is recovered more generally when $t_{10}/\omega\rightarrow 0$.

The first three terms corresponds to the diagonal terms which are
independent of the time between the hard production point and the
rescattering point. The next three terms are oscillating interference
terms. The color factor $R=C_A/C_R$ is 1 for a gluon jet and $9/4$ for
a quark jet.

It is interesting to note that if the momentum transfer $\vq_1$ \/ in
the scattering vanishes, then $\vec{B}_1=0$, $\vec{C}=\vec{H}$, and
$\omega_1=\omega_0$ and the above distribution reduces to the zero
scattering case in the previous section. This is a manifestation of
gauge invariance requiring a finite $\vq_1$ to generate a color dipole
that can modify the hard radiation pattern.

It is clear that the classical incoherent parton cascade limit
corresponds to assuming that all the oscillating interference term
average to zero.  In that case, the final gluon distribution is a sum
of the initial hard radiation plus an isolated Gunion-Bertsch
bremsstrahlung plus a rescattering ``cascade'' contribution reflecting
the modification of the final transverse momentum spectrum due to
rescattering of a gluon produced at $t_0$.  That latter term broadens
the transverse distribution of the emitted gluons.

\subsection{Relative Conditional Probability for $n_s=1$}

The magnitude of the interference effects in eq.~(\ref{dn10}) is controlled
by dimensionless variables
\begin{equation}
\kappa\equiv{\vk_\perp^{\; 2}\over \mu^2}\; ,\quad
\xi\equiv{\lambda \mu^2 \over 2\omega}={\omega_{BH} \over \omega}\;, \quad 
\kappa_1\equiv{(\vk-\vq_1)^2_\perp\over \mu^2}\;, \quad 
\theta_1\equiv{\vq_{1 \perp}^{\; 2}\over \mu^2} \; .
\end{equation}
Note that this averaged interference
cosine is  small when $\xi \kappa_i > 1$.
This is the kinematic region where the relevant formation time,
$\tau_i= 1/\omega_i=2\omega/(\vk_\perp-\vq_{i\perp})^2$,  is smaller
than the mean free path. For a fixed energy $\omega$ this restricts the angle 
of induced emission radiation to exceed 
$\theta_{min}=1/\surd(\omega \lambda)$. Below $\theta_{min}$ complete
destructive interference cancels all but the hard production amplitude.

The time averaged relative conditional
distribution for fixed
$\vq_{1 \perp}$ can be written as
\begin{eqnarray}
R^{(1)}_g(\vk,\vq_{1\perp})  &=& R^{(0)}_g(\vk)
{\cal F}^{(1)}(\vk_\perp,\vq_{1\perp},\xi) \; ,
\end{eqnarray}
where $R^{(0)}_g$ is given by eq.~(\ref{rhog0})
and the modification factor due to  $n_s=1$ rescattering is
given by 
\begin{eqnarray}
{\cal F}^{(1)} &=& \frac{\kappa^2}{\tilde\kappa^2}
+R\left(\frac{\kappa^2}{\tilde\kappa^2}+ 2 \frac{\kappa\kappa_1}
{\tilde\kappa_1^2}- 2\frac{\kappa\Delta \kappa}{\tilde\kappa
\tilde\kappa_1} \right)
-R\left(\frac{\kappa^2}{\tilde\kappa^2}-
\frac{\kappa\Delta \kappa}{\tilde\kappa \tilde\kappa_1} \right)
\frac{1}{1+\xi^2\tilde\kappa^2} \nonumber\\[.5ex]
&\;&\; -R\left(\frac{\kappa\Delta \kappa}{\tilde\kappa 
\tilde\kappa_1}\right)\frac{1}{1+\xi^2(\tilde\kappa-\tilde\kappa_1)^2} 
+2 R\left(\frac{\kappa\Delta \kappa}{\tilde\kappa \tilde\kappa_1}
- \frac{\kappa\kappa_1}{\tilde\kappa_1^2}\right)
\frac{1}{1+\xi^2 \tilde\kappa_1^2} \nonumber\\[.5ex]
&=&\frac{\kappa^2}{\tilde\kappa^2}+R\; \xi^2 \left\{\frac{\kappa^2}
{1+\xi^2\tilde\kappa^2}+ 2\frac{\kappa\kappa_1}{1+
\xi^2 \tilde\kappa_1^2} \right.\nonumber \\[.5ex]
&&- \left.\left(\frac{\kappa\Delta \kappa}{\tilde\kappa \tilde\kappa_1}
\right)\left(\frac{2 \tilde\kappa_1^2}{
1+\xi^2\tilde\kappa_1^2}-\frac{(\tilde\kappa-\tilde\kappa_1)^2}
{1+\xi^2(\tilde\kappa-\tilde\kappa_1)^2}
+\frac{\tilde\kappa^2}{1+\xi^2\tilde\kappa^2}
\right)\right\} \; ,\label{F1}
\end{eqnarray}
where $\Delta\kappa = \kappa - \vk_\perp\cdot\vq_{1 \perp}/\mu^2=(
\kappa^2+\kappa_1^2-\theta^2_1)/2$. 
The tilde denotes $\tilde \kappa_i \equiv \kappa_i +\epsilon^2$,
that results when a plasmon mass shift, $\omega_{pl}^2=
\epsilon^2\mu^2$, is introduced 
in all  propagators.

Eq.~(\ref{F1}) shows that for finite $\xi$, 
the  $\vk_\perp=\vq_{1\perp}$
singularity of free space perturbation theory eq.~(\ref{BGrad}) 
is automatically regulated by the interference terms. 
For finite $\xi$ therefore we can safely take the
$\epsilon=0$ limit  and write
\begin{eqnarray}
{\cal F}^{(1)} &=& 1+R \; \xi^2 \left\{ \frac{\kappa^2}{1+\xi^2\kappa^2}
+ 2\frac{\kappa\kappa_1}{1+\xi^2 \kappa_1^2} \right.\nonumber \\[.5ex]
&\;&\;  - \left.\left(\frac{\Delta \kappa}{\kappa_1}\right)
\left(\frac{2 \kappa_1^2}{1+\xi^2\kappa_1^2}-\frac{(\kappa-\kappa_1)^2}
{1+\xi^2(\kappa-\kappa_1)^2}+\frac{\kappa^2}{1+\xi^2\kappa^2}
\right)\right\} \;. \label{F10}
\end{eqnarray}

However, in the {\em formal} classical
parton cascade limit with $\xi\rightarrow \infty\;$ ($\omega\gg\omega_{BH}$)
the  apparent singularity is exposed unless $\epsilon\ne 0$: 
\begin{eqnarray}
{\cal F}^{(1)}(\vk_\perp,\vq_{1\perp},\xi=\infty)
 &=&\frac{\kappa^2}{\tilde\kappa^2}(1+R)
+2R\frac{\kappa}{\tilde\kappa_1}\left( \frac{\kappa_1}{\tilde\kappa_1}
-\frac{\Delta\kappa}{\tilde{\kappa}}\right) \label{f1eps} \nonumber \\[.5ex]
&\stackrel{\epsilon\rightarrow 0}{\longrightarrow}
& \left({\vec H}^{\;2}+R\left ({\vec B}_1^2 +{\vec C}_1^2 \right)\right)/
{\vec H}^2
 \label{f1inc} \; .
\end{eqnarray} 
To compare the full quantum parton
cascade modification factor eq.~(\ref{F10}) to the incoherent
classical parton cascade result, we regulated the singular
classical limit with $\epsilon=1$
\begin{equation}
{{\cal F}^{(1)}_{cl}(\vk_\perp,\vq_{1\perp}})
\equiv \left.{{\cal F}^{(1)}_{cl}(\vk_\perp,\vq_{1\perp}},\infty)
\right|_{\epsilon=1}
\; .\label{f1clas}\end{equation}

In the opposite $\xi\rightarrow 0$ limit, on the other
hand, we see that  the factorization theorem,
 ${\cal F}^{(1)}\rightarrow 1$, is recovered.

Another interesting limit is $k_\perp \gg q_{1\perp}\sim \mu$.
In that kinematic range, the induced GB bremsstrahlung can be ignored, 
(${\vec B}_1^2\approx 0$)
but the cascade term survives (${\vec C}_1^2\rightarrow{\vec H}^2$). 
Therefore, in the 
$\mu\ll k_\perp < \omega$ kinematic regime
\begin{eqnarray}
{\cal F}^{(1)}(\vk_\perp,\vq_{1\perp} )
&\stackrel{\vk_{\perp}^{\;2} \gg \mu^2}{\longrightarrow} & 
 1+R \; .\label{f1asm}
\end{eqnarray}
This enhancement of the high $k_\perp$ tail arises because
rescattering of the gluon is enhanced by a color factor $R=C_A/C_R$
relative to the jet. As we shall see, this is a special case of 
the general asymptotic $k_\perp^2 \gg n_s \mu^2$ form
in the case of $n_s$ collisions:
\begin{eqnarray}
{\cal F}^{(n_s)}(\vk_\perp,\vq_{1\perp},\cdots \vq_{n_s \perp} )
&\stackrel{\vk_{\perp}^{\;2} \gg n_s \mu^2}{\longrightarrow} & 
 (1+R)^{n_s} \; .\label{fnasm}
\end{eqnarray}

Although it is  possible to integrate over the momentum
transfers analytically, the result even in the case of one scattering
center is not instructive.  Therefore, we average ${\cal F}^{(1)}$
over $\vq_{1\perp}$ numerically for
different values of $\xi=\omega_{BH}/\omega$: 
\begin{equation} 
F^{(1)}(\kappa,\xi)= \langle\, {\cal F}^{(1)}\, \rangle
 = \int_0^{s/4\mu^2}\frac{d\theta_1}{(\theta_1+1)^2} 
\int_0^{2\pi}\frac{d\phi_1}{2\pi}
\, {\cal F}^{(1)}( \kappa, \theta_1, \phi_1,\xi)
 \; .
\end{equation} 
Note that we keep the large $s/4\mu^2\approx 6E_{jet}T/4\mu^2$ 
upper bound in the
integration finite because the kinematic boundaries include regimes
where $E^2\ge \omega^2 \ge k_\perp^2  \ge s$.

In Fig. 4a the three components of the
classical parton cascade envelop ($\xi=\infty$) for an initial quark jet
are  shown for the case $\mu=0.5$ GeV as a function
of the logarithmic ``angle'' $\log_{10}(\kappa)$. 
The Hard factorization term corresponds to
the unit curve. The GB induced radiation component dominates at small
$\kappa$ and the cascade contribution is only important at large $\kappa$.
The bump at $\kappa\sim 5$ is the remnant of the
logarithmic singularity exposed in the $\xi=\infty$
limit, but regulated here with a plasmon
mass $\omega_{pl}=\mu=0.5$~GeV.

Fig. 4b shows  the ensemble averaged
quantum modification factor, $F^{(1)}$, 
with $\mu=0.5$ GeV,$\;\;\lambda=1$ fm/c ( $\omega_{BH}\approx 0.63$ GeV)
for seven different gluon energies $\omega$.
These correspond to $\xi=0.013, \; 0.031, \; 0.063, \; 0.13,
\; 0.31, \; 0.63,$ and $1.3$
that cover an energy range $0.5\; {\mbox{GeV}}<\omega<50\; {\mbox{GeV}}$ . 
In each case, the curves are cut off at the kinematic bound 
$$\kappa_{max}=\frac{\omega^2}{\mu^2}=\frac{(\lambda\mu)^2}{4 \xi^2}\; ,$$
which corresponds to right angle radiation relative to the jet axis.
Here $(\lambda\mu)^2/4\approx 1.5\,$.
The curves illustrate well how the total destructive
interference (LPM effect) for $\kappa\xi < 1$ reduces $F^{(1)}$ to unity.
In particular, the induced Gunion-Bertsch radiation component
in Fig. 4a is almost completely suppressed due to interference with the hard
production amplitude. Only the cascade term modifies the spectrum considerably
at large $k_\perp$.
The onset of significant medium modification 
can be seen to follow qualitatively the point $\kappa_c= 1/\xi$.

\begin{center}
\vspace*{9.0cm}
\includegraphics{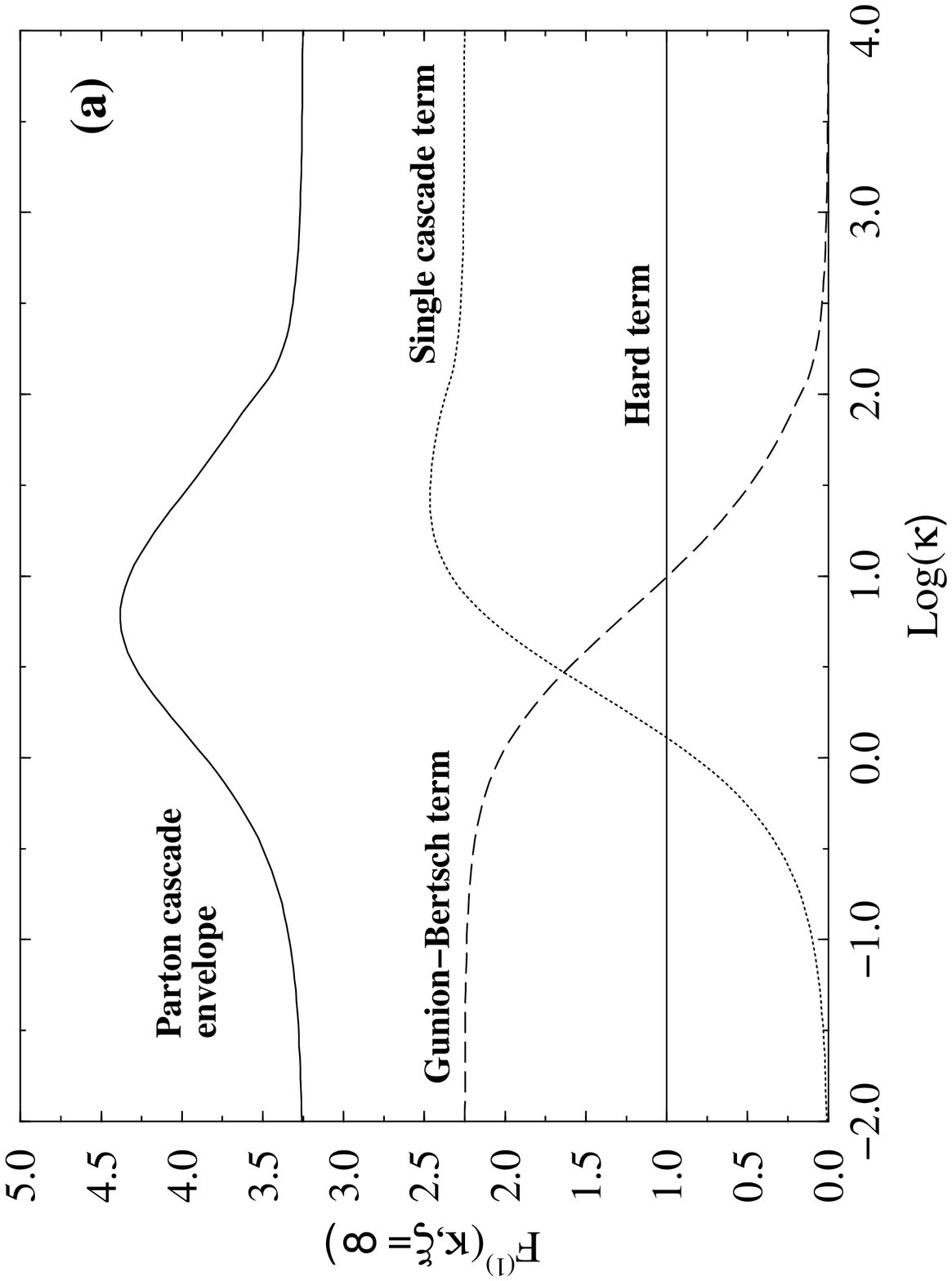}
\includegraphics{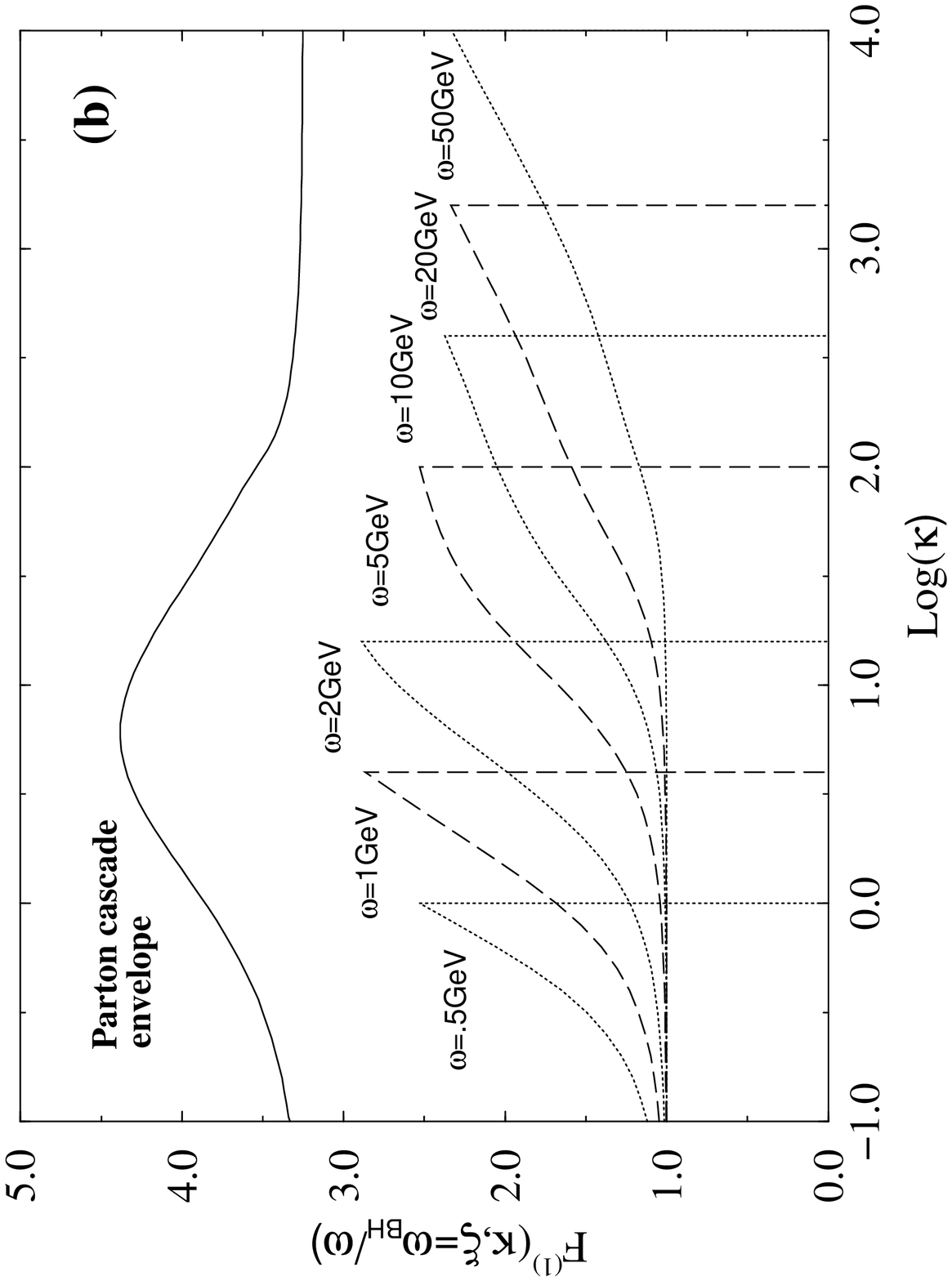}
\vskip -45pt
\begin{minipage}[t]{15. cm}
{\small {\bf Fig.~4.}  (a) The normalized time-independent envelopes and
separate contributions to the conditional
probability distribution of gluons associated with 
a single rescattering $n_s=1$ of quark jet (${\mbox E}_{\mbox
{jet}}=50{\mbox {GeV}}$) is shown
as a function of the logarithmic ``angle'', 
$\log_{10} k_\perp^2/\mu^2$ with $\mu=0.5$ GeV.
These curves correspond to the incoherent parton cascade limit.
(b) For finite $t_{10}=1$ fm/c case destructive interference
limits the corrections to the hard self-quenched
distribution to higher angles as shown for different gluon energies.
}
\end{minipage}
\end{center}
\vskip 4truemm

It is convenient to define in general the reduced modification factor
$f^{(n_s)}(\kappa,\xi)$ through
\begin{equation}
F^{(n_s)}(\kappa,\xi)\equiv 
\left\{1+ \left( (1+R)^{n_s} -1\right) 
\; f^{(n_s)}(\kappa, \xi) \right\}
\; \theta\left(\frac{(\lambda\mu)^2}{4\xi^2}-\kappa\right)  \; ,
\label{fapprox}
\end{equation} 
which isolates the kinematic boundary and the
 asymptotic color scaling factor.
Note that the incoherent classical parton cascade envelope
is not  reached for any reasonable values of $\lambda \mu$ 
due the   finite kinematic boundaries.

\section{Case of Two Scattering Centers}

The relative conditional probability  $R_g^{(2)}(\vk\,)$
can be obtained through the averaging of the square of the matrix
element ${\cal M}_2$ on the scattering potentials in the 
$\vq_1$ and $\vq_2$ space.

In Appendix A we listed the matrix elements connected to the seven
 planar diagrams, contributing to the soft gluon radiation. 
The sum of the matrix elements can be written as
\begin{eqnarray}
{\cal M }_2   
&=& -2 i g_s e^{i t_0 \omega_0} \times \vec{\epsilon}_\perp
\cdot \left\{ \vec{H}a_2a_1c + \vec{B}_1  e^{i t_{10} \omega_0}
a_2 \left [ c, a_1 \right ]  \right. \nonumber \\[1.5ex]
 && +\, \vec{B}_2 e^{i t_{20} \omega_0}
 \left [ c, a_2 \right ] a_1 +  
\vec{B}_{2(12)}e^{i (t_{20} \omega_{0}-t_{21}\omega_2 )} 
\left [ \left [ c, a_2 \right ] , a_1 \right ]   \nonumber \\[1.5ex]
&&+\, \vec{C}_1  e^{-i t_{10} \omega_{10}}
a_2 \left [ c, a_1 \right ] +\, \vec{C}_2 e^{-i t_{20} \omega_{20}}
a_1 \left [ c, a_2 \right ]  \nonumber \\[1.5ex]
&&  \left. + \, \vec{C}_{(12)} 
 e^{i( t_{20} \omega_{0}- t_{21}\omega_2 - t_{10}\omega_{(12)})}
\left [ \left [ c, a_2 \right ] , a_1 \right ] \,
\right\}  \; . \label{n2} 
\end{eqnarray} 
In the limit where all $t_i-t_j =0, \ 
t_i > t_j, \ i, j = 1,2$ one can easily see that all the terms but 
the ``hard'' scattering cancel 
\begin{eqnarray}
{\cal M }_2 &=& -2 i g_s e^{i t_{0} \omega_0}\,\vec{\epsilon}_\perp 
\cdot\vec{H}  a_2 a_1 c \;. 
\end{eqnarray}  
The GB $t_0 \rightarrow -\infty$ limit and the connection to the 
BDMPS ~\cite{BDMPS} result we discuss in Appendix B.

After calculating the appropriate color factors in
eqs.~(\ref{colo1})-(\ref{colo2}) 
we obtain 
\begin{eqnarray}
\tr \left ( {\cal M}_2 {\cal M}^\dagger_2 \right )
& = &C^3_RD_R \times \left\{
\vec{H}^2 + R \left( \vec{B}^2_1+\vec{B}^2_2+\vec{C}^2_1+\vec{C}^2_2
\right ) \right. \nonumber \\ 
&& \left. + R^2\left ( \vec{B}^2_{2(12)}+ \vec{C}^2_{(12)} \right ) 
\; \right\} +{\rm Interference \ terms} \; . \label{2cos}
\end{eqnarray}
(See eq.~(\ref{neq2})   for the full result.)
We can isolate  the color prefactor together with the leading parabolic
dependence $1/k^2_\perp \sim R^{(0)}_g$ in the form 
\begin{equation}
R^{(2)}_g(\vk,\vq_{1\perp})  = R^{(0)}_g(\vk)
{\cal F}^{(2)}(\vk_\perp,\vq_{1\perp},\vq_{2\perp},\xi_1,\xi_2) \; .
\end{equation}
We note that when the interference terms drop at asymptotic 
$k_\perp^2 \gg 2 \mu^2$ 
the enhancement factor scales as in eq.~(\ref{fnasm})
\begin{eqnarray}
{\cal F}^{(2)}(\vk_\perp,\vq_{1\perp}, \vq_{2 \perp} )
&\stackrel{\vk_{\perp}^{\;2} \gg 2\mu^2}{\longrightarrow} & 
 (1+R)^{2} \; .\label{f2asm}
\end{eqnarray}

The longitudinal  ensemble averaging can be performed in eq.~(\ref{2cos}) 
over the times $t_{10}=t_1-t_0\;, \; t_{21}=t_2-t_1$   as indicated in 
eqs.~(\ref{zdis},\ref{expff}). For the more general case $ \lambda_1 \neq 
\lambda_2$ we find 
\begin{equation}
\left \langle \; \cos (\omega_\alpha t_{10} + \omega_\beta t_{21} )
\; \right \rangle  = { 1 - \lambda_1 \lambda_2 \omega_\alpha \omega_\beta 
\over (1+\lambda^2_1\omega^2_\alpha) (1+\lambda^2_2\omega^2_\beta)} \; .
\end{equation}

The integration over the  momentum transfer  due to the rescatterings 
from the static centers is most readily performed in terms of the dimensionless 
variables
\begin{equation}
\kappa={k_\perp^2\over \mu^2},\quad \theta_i={q_{i\perp}^2\over \mu^2},
\quad \xi_i={(t_i-t_{i-1})\mu^2 \over 2\omega}, \quad i=1,2 \; ,
\end{equation}
leading to an averaged quantum correction factor 
\begin{eqnarray} 
F^{(2)}(\kappa,\xi_1,\xi_2)&=& \left \langle \; 
{\cal F}^{(2)}(\kappa,\vq_{1\perp},\vq_{2\perp},\xi_1,\xi_2) \;
\right \rangle =\; \nonumber \\[.5ex]
& =& 
\int_0^{s/4\mu^2}\frac{d\theta_1}{(\theta_1+1)^2} 
\frac{d\theta_2}{(\theta_2+1)^2}
\int_0^{2\pi}\frac{d\phi_1}{2\pi}\frac{d\phi_2}{2\pi} 
\,{\cal F}^{(2)}
 \; .
\end{eqnarray}

On Fig.~5a we plot the contributions to  ${\cal F}^{(2)}_{cl}$ -- the incoherent
parton cascade limit where all the exposed singularities in the propagators
have been regulated by the Debye screening mass $\mu$.  
At small transverse momenta of the radiated gluon the quantum modification is dominated by 
the ``Gunion-Bertsch terms''. In the limit of large $k^2_\perp / \mu^2$ the  only terms 
that survive apart from the hard scattering are the ``cascade terms''. This represents the 
expected behavior of $R^{(2)}_g(\vk\,)$ in the Bethe-Heitler limit.
It is important to notice that simple picture of the process is only possible
in the incoherent case where one easily decouples production from rescattering
and the separate terms acquire physical interpretation.

Fig.~5b shows  ${\cal F}^{(2)}$ plotted for the same set of gluon energies 
$\omega$
as in Fig.~4b (and the same values of $\xi_i,\;\; i=1,2$). The picture
of classical parton cascading is modified by the presence of  interference
terms that control behavior of the probability distribution in the 
intermediate LPM region. Due to the destructive interference one observes 
a smooth transition from the ``hard term'' at small $k_\perp^2$ 
(i.e. the factorization limit) to  the Bethe-Heitler limit. 
\begin{center}
\vspace*{9.0cm}
\includegraphics{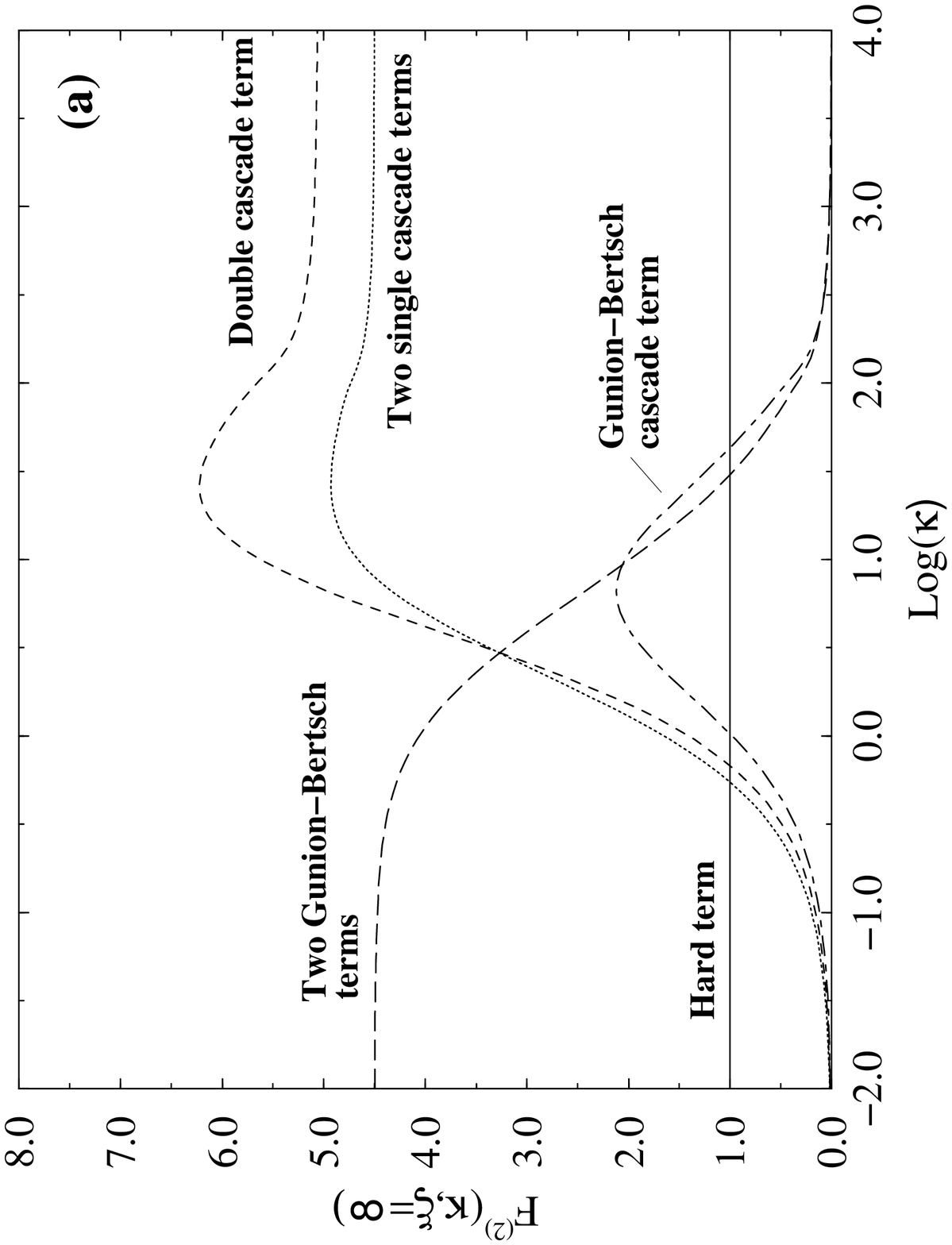}
\includegraphics{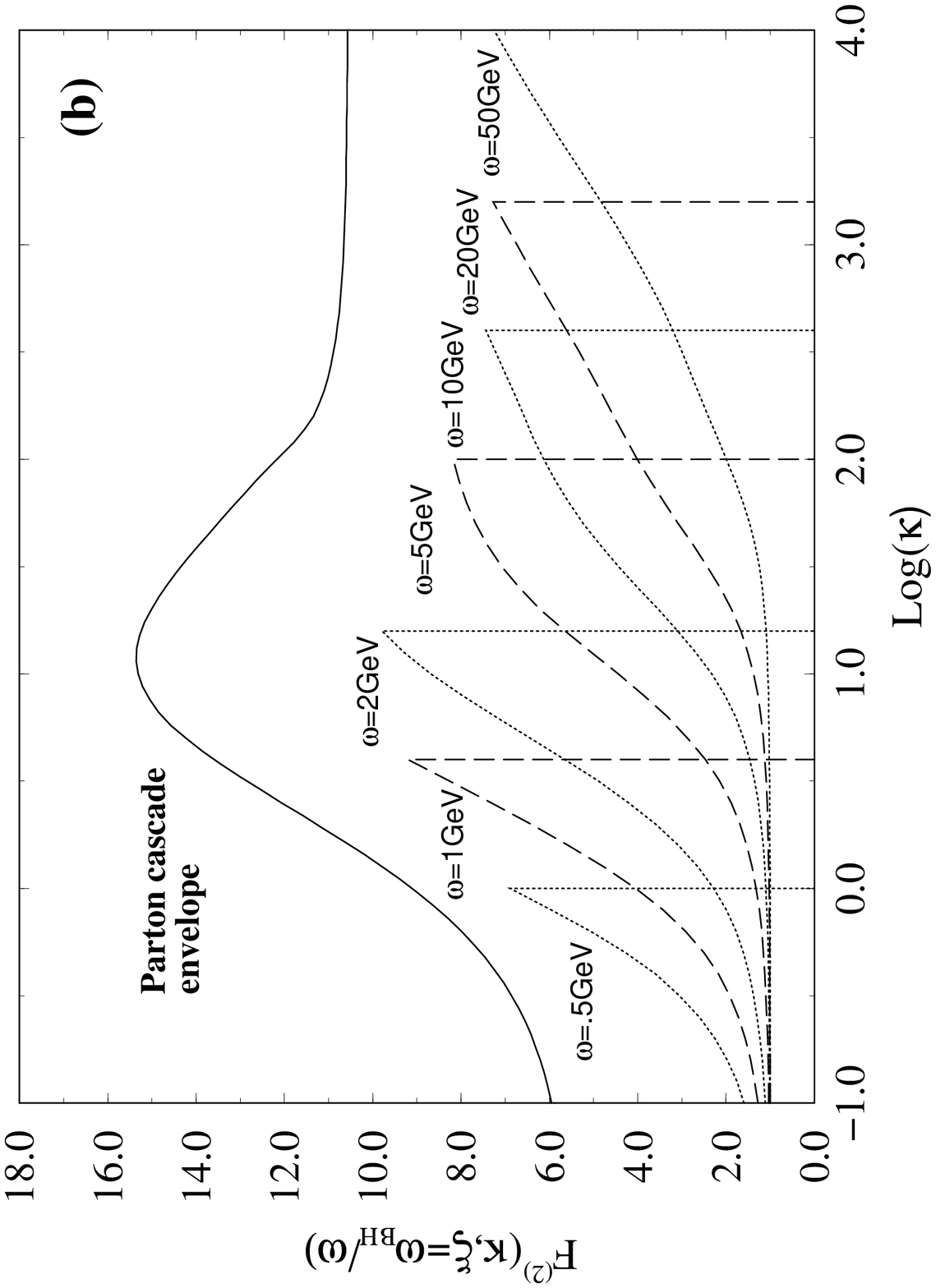}
\vskip -45pt
\begin{minipage}[t]{15.0 cm}
{\small {\bf Fig.~5.}  (a) The structure of the regulated ($\mu=0.5\; $GeV) 
contributions to the $\xi \rightarrow \infty$ limit
(quark jet of ${\mbox E}_{\mbox{jet}}=50\;{\mbox {GeV}}$, $n_s=2$).
(b) Characteristic behavior of $F^{(2)}$ for 
$\xi=0.013, \; 0.031, \; 0.063, \; 0.13,
\; 0.31, \; 0.63$ and $1.3\;$.
}
\end{minipage}
\end{center}
\vskip 4truemm
It is remarkable that the detailed angular profile of the 
probability distribution is very similar (apart from the
explicit rescaling factor) to the one in Fig. 4b.

\section{Power Law Scaling in $n_s$ and Estimated Energy Loss}

The amplitudes in  the case of $n_s=3$ are tabulated in Appendix C.
The main novelty at this order is the appearance of an
interesting ``color wheel'' diagram derived in Appendix D. The amplitude,
${\cal M}_3$ is rearranged in eq. (\ref{n3}) 
in terms of the hard, GB, and cascade
amplitudes. The diagonal (classical parton cascade)
terms in $|{\cal M}_3|^2$ with color factors 
(\ref{diag3}) involve (1) the usual ${\vec H}^{2}=1/k_\perp^2$ with a 
relative color weight 1, (2) three simple GB induced
radiation terms and three one scattering gluon cascade
terms each with color weight of $R=C_A/C_R$, (4) three double gluon cascade
terms and four GB-cascade terms with a weight $R^2$, and (5) 
 one triple gluon cascade term with a weight $R^3$.
At small $k_\perp$, only the three simple GB (${\vec B}^{2}\rightarrow 
{\vec H}^{2}$)  terms contribute since
the finite cascade terms are negligible relative to the divergent
hard and GB terms. In this small $k_\perp$ region therefore
the ${\vec H}^{2}$ term is simply amplified by a factor $1+3R$ as illustrated
 in Fig.~6a. For large angles, on the other hand, all the GB and GB-cascade
terms vanish while the cascade terms ${\vec C}^{2}\rightarrow 
{\vec H}^{2}$. Therefore, in that region the hard radiation is amplified by a
factor $(1+R)^3$. In between, near $k_\perp\sim \mu$
all terms contribute and a bump that is sensitive to the screening
scale $\mu$. Fig.~6a summarizes therefore the
general scaling with $n_s$ of the classical parton cascade angular
distribution. The power law enhancement $(1+R)^{n_s}$ at large angles
is the most important nonlinear feature since the bump near $\kappa \sim 1$
is killed by destructive interference shown on Fig. 6b.

\begin{center}
\vspace*{9.0cm}
\includegraphics{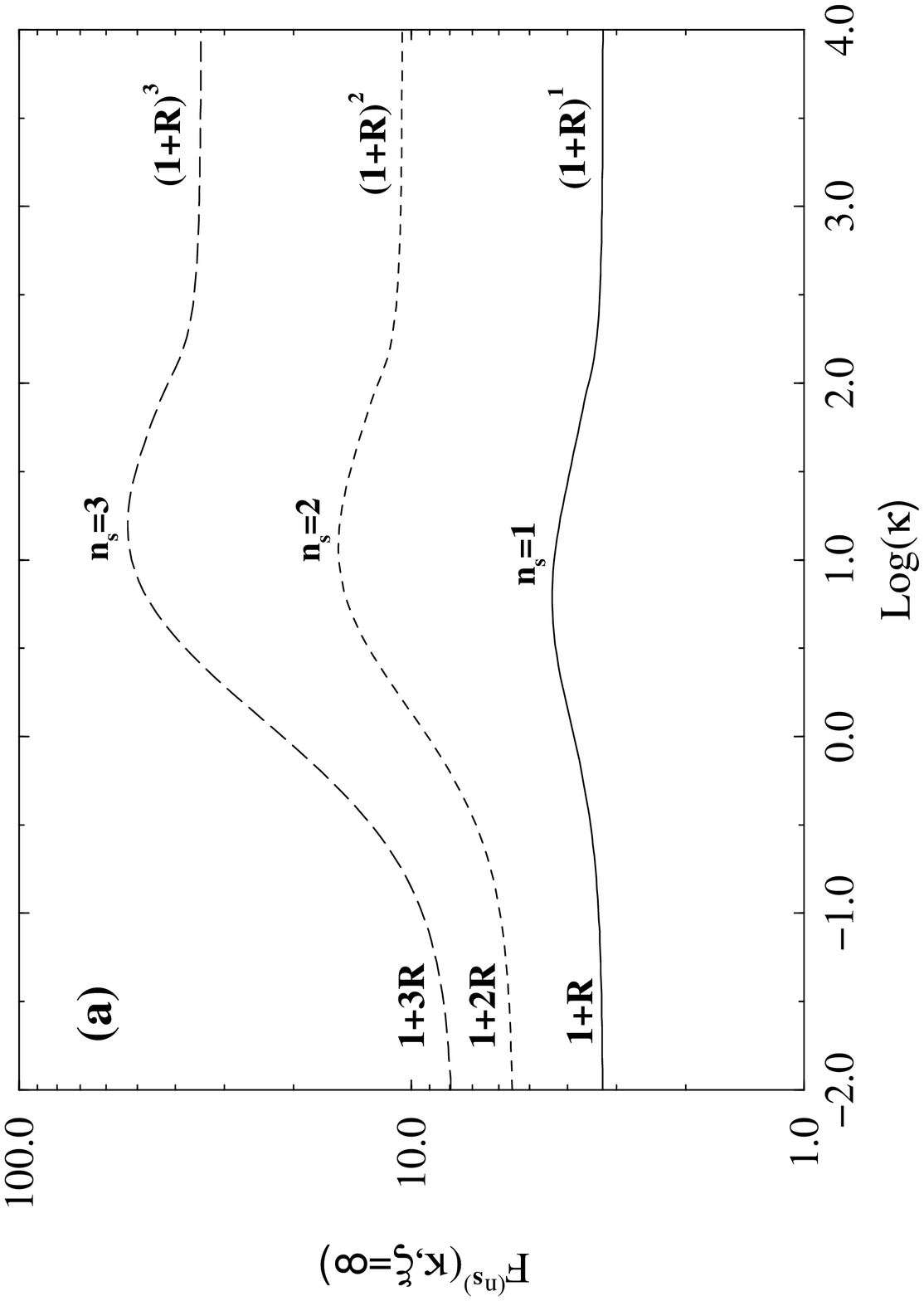}
\includegraphics{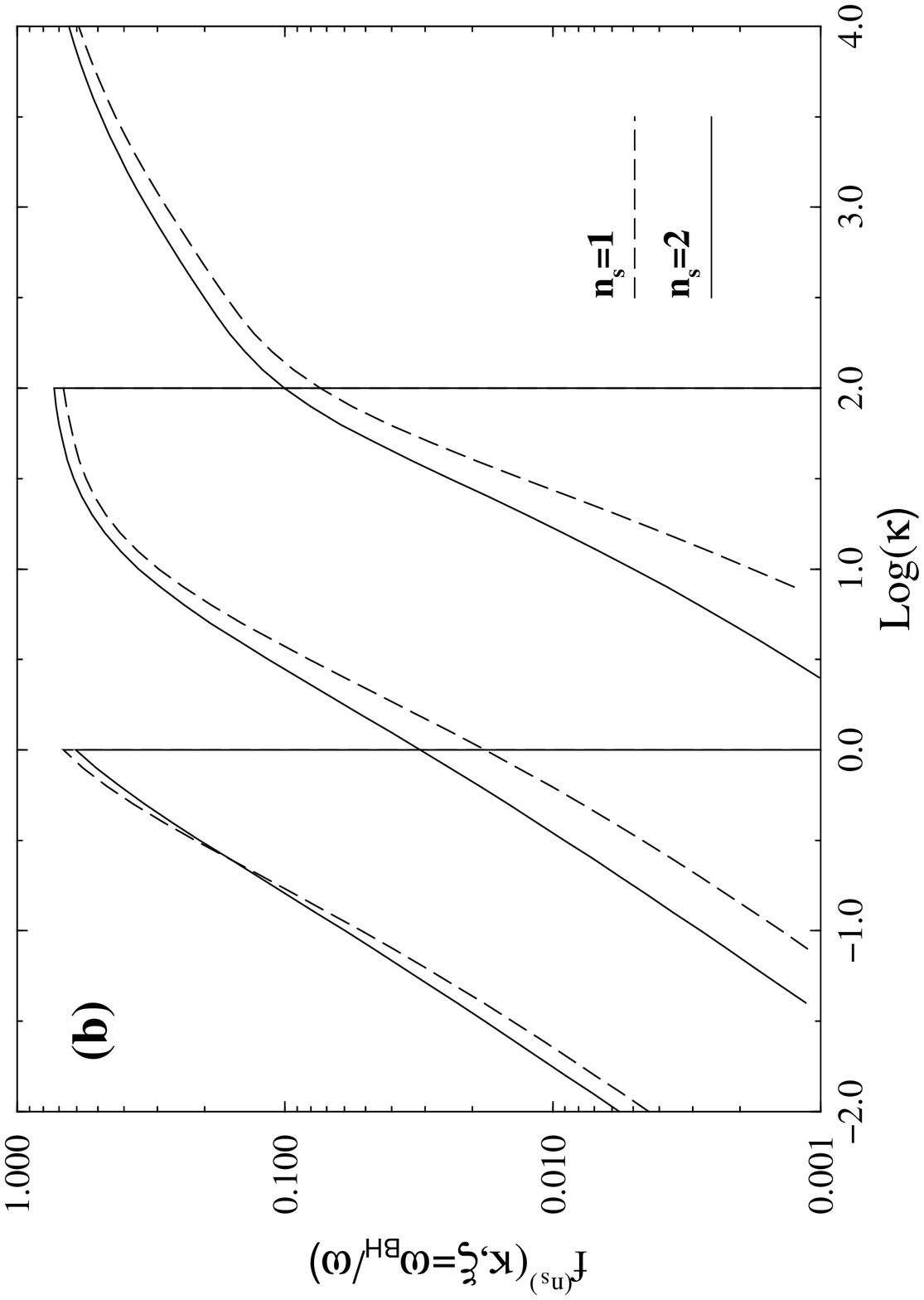}
\vskip -45pt
\begin{minipage}[t]{15.cm}
{\small {\bf Fig.~6.}
(a) The classical parton cascade limit ($\xi \rightarrow \infty$)
angular enhancement factor 
from eq.~(\ref{fapprox}) for
 a quark jet (${\mbox E}_{\mbox{jet}}=50\;{\mbox {GeV}}$, 
 $\mu=0.5$ GeV) vs $\log_{10}(k_\perp^2/\mu^2)$. 
(b) The reduced modification
factor $f^{(n_s)}$  including kinematic cut-offs for  
$n_s=1,2\;$ shows an approximate $n_s$ independence.}
\end{minipage}
\end{center}

In the general case, the small $k_\perp$ enhancement, $1+ n_s R$,
expected in the classical cascade 
limit also is cancelled by destructive interference as seen in Figs.
4 and 5. Here we note a remarkable simple scaling between the $n_s=1,2$
cases. In Fig. 6b the reduced modification factors
$f^{(n_s)}(\kappa,\xi)$ extracted from $F^{(n_s)}(\kappa,\xi)$
in eq.~(\ref{fapprox}) are shown. The reduced factors are
nearly identical in spite of the much more complex functional
form of the $R^{(2)}_g$ than $R^{(1)}_g$.   

The approximate scaling from $n_s=1$ to $n_s=2$  can be summarized as follows:
\begin{equation}
R^{(n_s)}_g(\omega,\vk_\perp)\approx R^{(0)}_g(k_\perp) 
\left\{1+ \left( (1+R)^{n_s} -1\right) 
\; f^{(1)}(\kappa, \xi) \right\} \; .
\label{fscale}
\end{equation} 
The generic analytic structure of the reduced
modification factor can be roughly approximated  by
$ f^{(1)}(\kappa,\xi)\sim \kappa^2\xi^2/(1+\kappa^2\xi^2)$,
which shows that the classical parton cascade result
is approached only for $\kappa> 1/\xi$,
 i.e., $k_\perp > \sqrt{\omega/\lambda}$
For higher $n_s$ the validity of this scaling  
will be tested in~\cite{GLVII}.

Here we extrapolate  eq.~(\ref{fscale}) to  $n_s=3$ to make a rough 
estimate of 
 the energy loss spectrum $dI/d\omega$ and
the fractional energy loss $\Delta E/E$ to $L\sim 3$ fm.
We emphasize that these estimates are  included in this paper
only to illustrate qualitative features. For quantitative
estimates, multi-gluon showers that
require a Monte Carlo analysis  must be 
considered as deferred to ref.~\cite{GLVII}.
The normalization factor $Z_{n_s} \gg  1$ 
for even modest $n_s \sim 3$ in eq.~(\ref{z})
signals a high probability of multi-gluon emission. This  is  
well known  in the $n_s=0$ case as emphasized in eqs.~(\ref{z0},\ref{multg}). For 
example, $Z_0 > 2$ from eq.~(\ref{z0}) for
 $E > 10\;$GeV while DLA gives, $Z_0\approx 3$. Eq.~(\ref{multg}) then
leads 
to  $\langle n_g \rangle\approx 1.2$ 
for this energy,
implying that two gluon final states must be taken into account.

\begin{center}
\vspace*{9.0cm}
\includegraphics{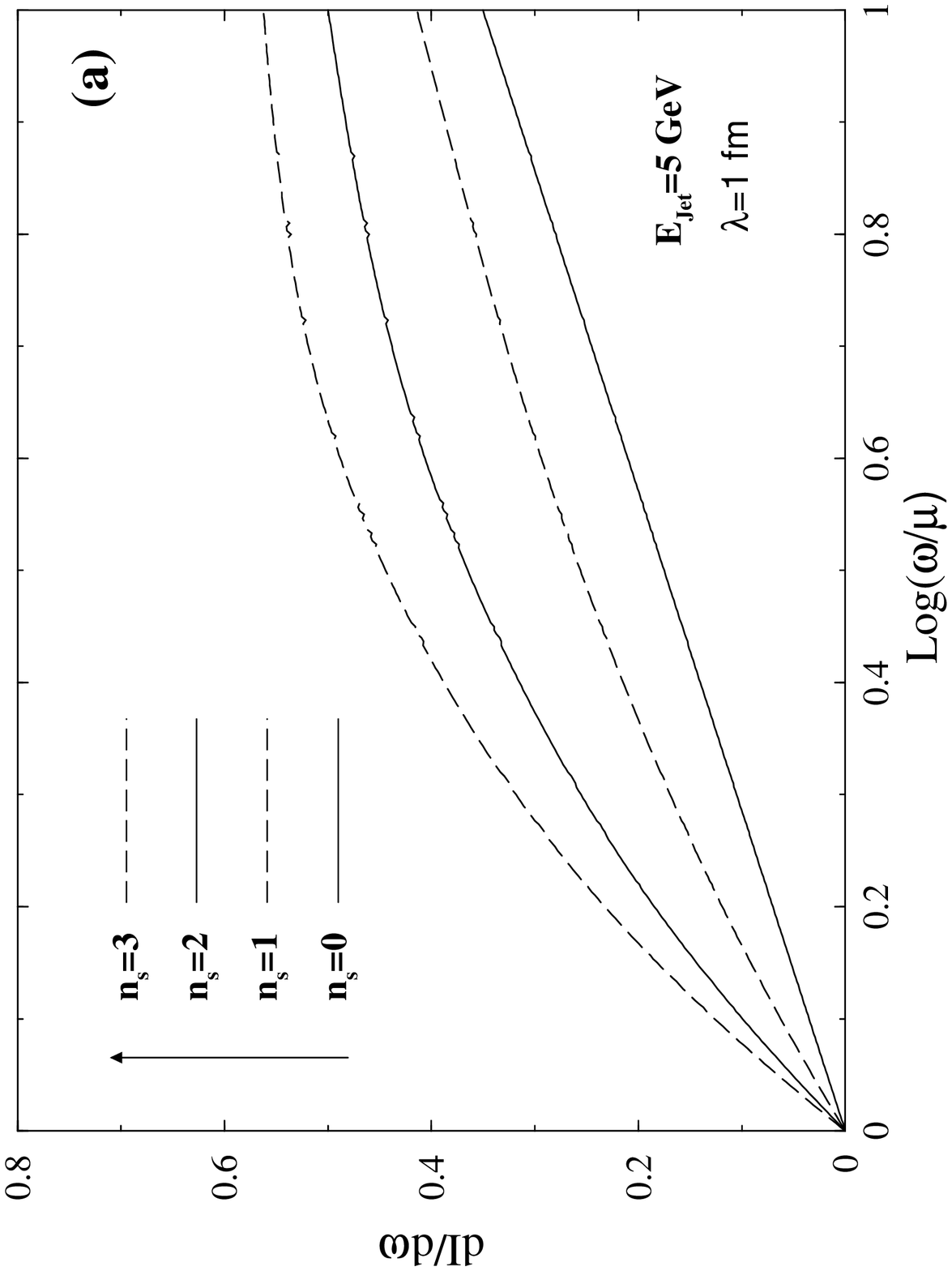}
\includegraphics{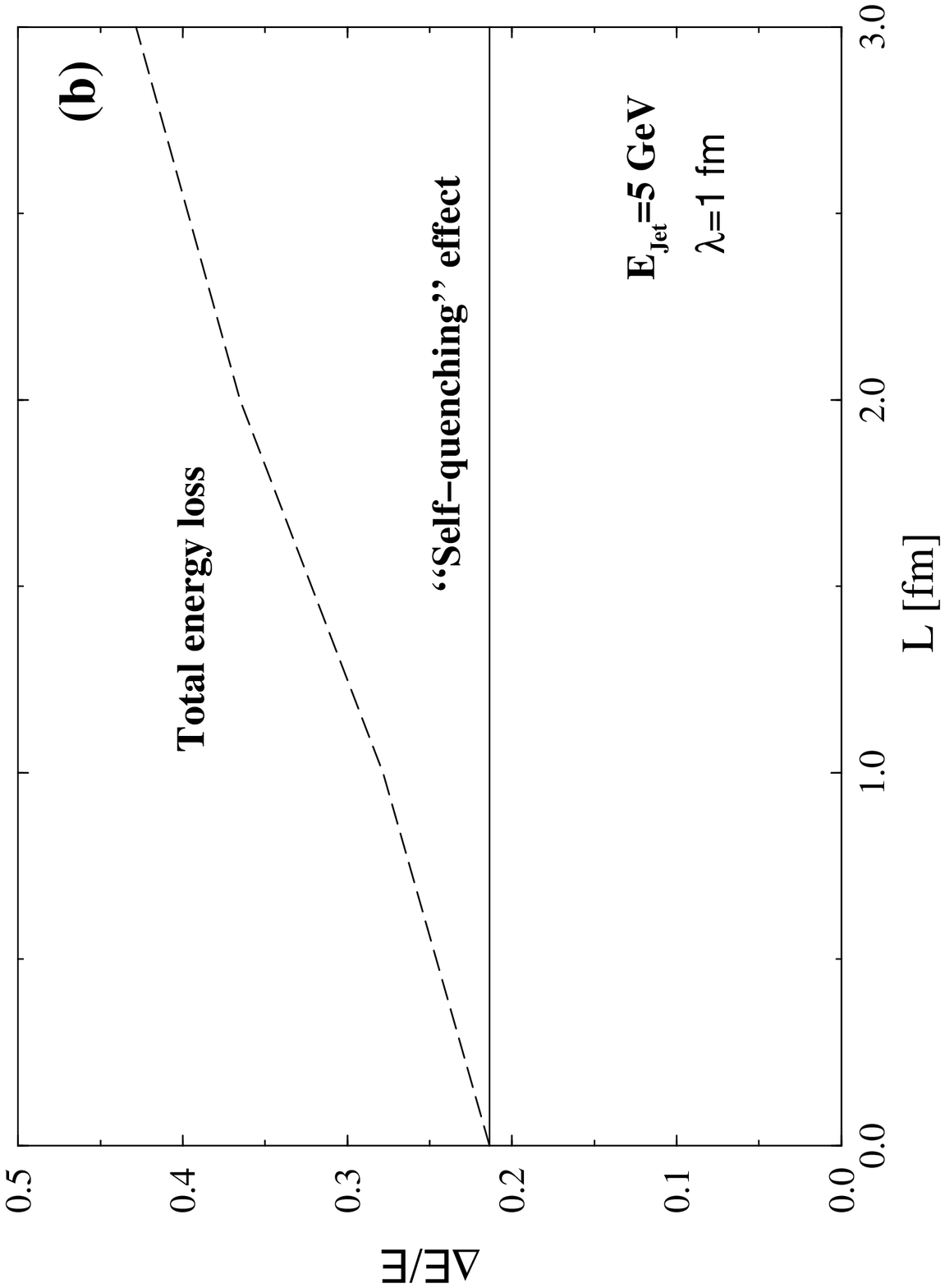}
\vskip -45pt
\begin{minipage}[t]{15.cm}
{\small {\bf Fig.~7.}
(a) The gluon radiation intensity $dI/dw$ for a 5 GeV quark jet ($n_s=0,1,2,3$) based on \protect{(\ref{fscale})} is illustrated.
(b) The fractional  energy loss is shown for the ``self-quenching''
and medium modified cases.}
\end{minipage}
\end{center} 

With the above caveats, Fig. 7 illustrates the qualitative
features of the angular
integrated $dI/d\omega$ and $\Delta E/E$ for a quark jet of energy
$5\;$GeV penetrating a thin ($n_s=1,2,3$)
plasma with $\mu=0.5$ GeV, $\lambda=1$ fm, $\alpha_s=0.3$.
As seen from eq.~(\ref{rhog0}), the intensity distribution
for the self-quenching ($n_s=0)$ case 
increases linearly as a function
of $\log(\omega/\mu)$  and the fractional energy loss due to self-quenching is 
proportional to $\log(E/\mu)$.

With increasing $n_s$, the approximate scaling in eq.~(\ref{fscale})
leads to:
$$\omega\frac{dN_g^{(n_s,n_g=1)}}{d^3\;\vec{k}}
 \approx \frac{1}{Z_{n_s}} \frac{C_R \alpha_s}{\pi^2}\frac{1}{k_\perp^2}
\left\{1 + ((1+R)^{n_s}-1) f^{(1)}(\kappa,\xi)\right\}\;,$$
\begin{equation}
Z_{n_s} \approx 1+ \frac{C_R \alpha_s}{\pi}
\int_\mu^E \frac{d\omega}{\omega}\int_{\mu^2}^{\omega^2}
\frac{dk_\perp^2}{k_\perp^2} \left\{1 + ((1+R)^{n_s}-1)
f^{(1)}\left(\frac{k_\perp^2}{\mu^2},\frac{\omega_{BH}}{\omega}\right )
\right\} 
\; . \label{rapp}
\end{equation}
Without the $Z$ factor, the gluon distribution would
scale approximately exponentially with plasma thickness
at high $k_\perp> \mu/\surd\xi$
due to the multiple collisions of the gluon in the target.
However, the normalization factor compensates this increase to a very large extent
as seen in Fig. 7. In fact, at this  single gluon level
of approximation, the energy loss is surprisingly small, $\sim 300$ MeV/fm,
and only grows approximately linearly
with increasing thickness $L=n_s\lambda$ in spite of the large modification
of the gluon angular distribution.

\section{Summary}

We investigated the angular distribution of radiated gluons
and the energy loss of hard jet in thin plasmas, where
the size of the plasma $L$ is comparable to the 
the mean free path, $\lambda$. The small number of
rescattering terms  made it possible to calculate analytically
the full matrix element within the eikonal approximation.
This paper established a systematic method for computation of all
relevant partonic matrix elements for higher $n_s$
which are needed as input to a numerical Monte-Carlo
study of the actual hadronic observables 
associated with jet quenching to be
presented in a subsequent paper \cite{GLVII}.  
We developed an efficient algorithm to
automate the cumbersome color algebra and keep track of the many canceling
phases involved in the non-abelian Landau-Pomeranchuk-Migdal effect.  
Detailed discussion for
the  $n_s=1,2,3$ cases were presented
to clarify the how the competing physical processes
interfere in  
different kinematic regions.
An approximate power law scaling of the 
angular broadening emerged with the increasing 
number of the rescattering centers. However,
the rapidly growing  wavefunction renormalization was shown
to limit the growth of final energy loss. For quantitative comparison
with data,  multigluon emission will be shown in
 GLV II\cite{GLVII} to dominate the jet quenching pattern
at the observable hadronic level.

\vspace{1.5cm}

\noindent{\large\bf Acknowledgments \\}

\noindent We thank Urs Wiedemann for extensive discussions related to
this problem.
This work was supported by the DOE Research Grant under Contract No.
De-FG-02-92ER-40764, partly by the US-Hungarian Joint Fund No.652
and OTKA No. T025579.

\newpage

\begin{appendix}

\vspace{3cm}

\section{Case of Two Scattering Centers}

In the approximation of a high energy incident parton and
soft gluon brems\-strah\-lung the seven planar diagrams in Fig.~8 
 can be treated in the eikonal approximation spinless limit. 

\begin{center}
\vspace*{17.7cm}
\includegraphics{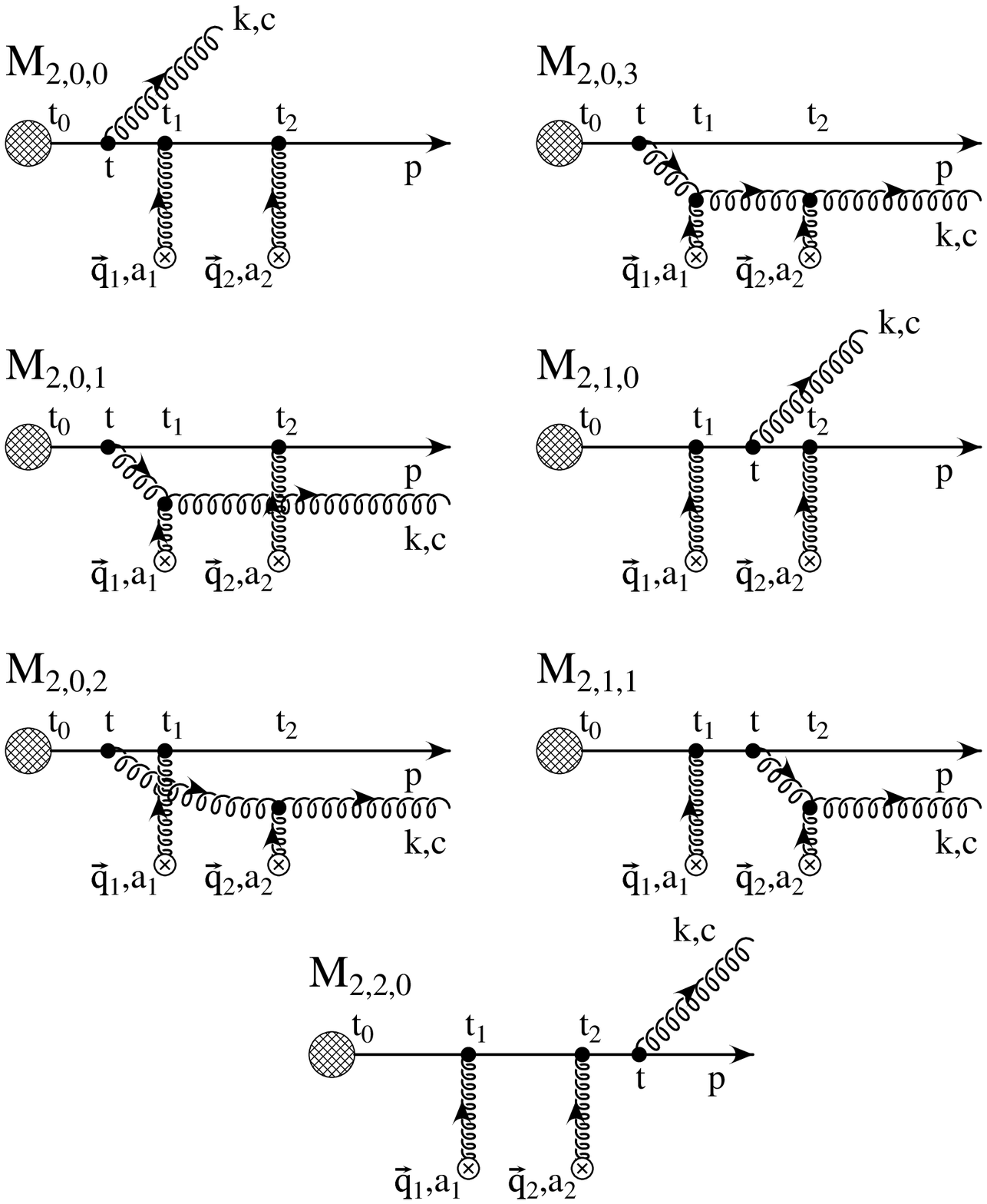}
\vskip -30pt
\begin{minipage}[t]{15.0cm}
{\small {\bf Fig.~8.}
The contributions to {\cal M}$_J \otimes$ {\cal M}$_2$  
soft gluon radiation amplitude in case of two scatterings.}
\end{minipage}
\end{center}
\vskip 4truemm

We denote 
by $t_0$ the time at which the particle is produced in the
medium and by $t_i, \ i=1, 2$ the times of the successive 
rescatterings. In the approximation when $\lambda \gg \mu^{-1}$
``backward'' scattering
can be neglected and the process is approximately
time ordered ~\cite{BDMPS}. We denote by
$\vk$ the momentum of the radiated gluon and by $\vq_i, \ 
i=1, 2$ are the momentum transfers. In QCD with each vertex we
associate a generator of the $SU(3)$ group. We denote
them by $c$ for the radiated gluon and $a_i, \ i=1, 2$ for the
momentum transfer vertices. The polarization of the radiated
gluon is determined by~$\vec{\epsilon}_\perp$.

The corresponding amplitudes can be obtained with the 
procedure described in eqs.~(\ref{kmask},\ref{phiml}-\ref{exact}):
\begin{eqnarray}
{\cal{M}}_{2,0,0} &=& 2 i g_s {\vec{\epsilon}_\perp \cdot
\vk_\perp \over k^2_\perp } 
(e^{i t_1 {k^2_\perp \over 2\omega}} 
-e^{i t_0 {k^2_\perp \over 2\omega}} ) a_2 a_1 c \; \;, \\[.5ex]
{\cal{M}}_{2,0,1} &=& 2 i g_s {\vec{\epsilon}_\perp \cdot
(\vk - \vq_1 )_\perp \over (k - q_1)^2_\perp } 
e^{i t_1 {k^2_\perp - (k - q_1)^2_\perp \over 2\omega}} \times \nonumber \\[.5ex]
 \qquad && \times (e^{i t_1 {(k - q_1)^2_\perp \over 2\omega}} 
-e^{i t_0 {(k - q_1)^2_\perp \over 2\omega}} ) 
 a_2 \left [ c, a_1 \right ] \; ,  \\[.5ex]
{\cal{M}}_{2,0,2} &=& 2 i g_s {\vec{\epsilon}_\perp \cdot
(\vk - \vq_2 )_\perp \over (k - q_2)^2_\perp } 
e^{i t_2 {k^2_\perp - (k - q_2)^2_\perp \over 2\omega}} \times \nonumber \\[.5ex]
 \qquad && \times (e^{i t_1 {(k - q_2)^2_\perp \over 2\omega}} 
-e^{i t_0 {(k - q_2)^2_\perp \over 2\omega}} ) 
a_1  \left [ c, a_2 \right ]  \; ,  \\[.5ex]
{\cal{M}}_{2,0,3} &=& 2 i g_s {\vec{\epsilon}_\perp \cdot
(\vk - \vq_1 -\vq_2 )_\perp \over (k - q_1 - q_2)^2_\perp } 
e^{i t_2 {k^2_\perp - (k - q_2)^2_\perp \over 2\omega}} 
e^{i t_1 {(k-q_2)^2_\perp - (k-q_1-q_2)^2_\perp \over 2\omega}} 
\times \nonumber \\[.5ex]
 \qquad && \times (e^{i t_1 {(k-q_1-q_2)^2_\perp \over 2\omega}} 
-e^{i t_0 {(k-q_1-q_2)^2_\perp \over 2\omega}} ) 
\left[ \left [ c, a_2 \right ], a_1\right] \;  \;, \\[.5ex] 
{\cal{M}}_{2,1,0} &=& 2 i g_s {\vec{\epsilon}_\perp \cdot
\vk_\perp \over k^2_\perp } 
(e^{i t_2 {k^2_\perp \over 2\omega}} 
-e^{i t_1 {k^2_\perp \over 2\omega}} ) a_2 c a_1 \; \;, \\[.5ex]
{\cal{M}}_{2,1,1} &=& 2 i g_s {\vec{\epsilon}_\perp \cdot
(\vk - \vq_2 )_\perp \over (k - q_2)^2_\perp } 
e^{i t_2 {k^2_\perp - (k - q_2)^2_\perp \over 2\omega}} \times \nonumber \\[.5ex]
 \qquad && \times (e^{i t_2 {(k - q_2)^2_\perp \over 2\omega}} 
-e^{i t_1 {(k - q_2)^2_\perp \over 2\omega}} ) 
\left [ c, a_2 \right ] a_1 \; \;,  \\[.5ex]
{\cal{M}}_{2,2,0} &=& 2 i g_s {\vec{\epsilon}_\perp \cdot
\vk_\perp \over k^2_\perp } 
( -e^{i t_2 {k^2_\perp \over 2\omega}} ) c a_2 a_1 \; \;.
\end{eqnarray}

After regrouping the sum of the above amplitudes  
five distinct color factors remain in eq.~(\ref{n2}). We label them as
follows:
\begin{eqnarray}
& {\cal C}_1=a_2a_1c\; , \quad {\cal C}_2=a_2 \left [ c , a_1 \right ]
\; , \quad {\cal C}_3= \left [ c , a_2 \right ] a_1 \; ,& \nonumber \\[.5ex]
& {\cal C}_4=a_1 \left [ c , a_2 \right ]
\; , \quad {\cal C}_5 =\left [  \left [ c , a_2 \right ] , a_1 
\right ] \; .&
\end{eqnarray} 
In the case of QCD the ``color'' and ``kinematic'' parts of
a diagram factor. Diagrammatic techniques have been developed to 
treat the color part alone~\cite{Cvit}.  

We denote by $C_R$ the Casimir of the representation of the incident parton. 
For $SU(N)$ following the standard normalization for the generators 
we have
\begin{equation}
C_F={N^2-1 \over 2N}, \qquad C_A=N.
\end{equation}
The trace of the color factors can be taken separately in the square of the
matrix element, yielding
\vskip 1pt
{\footnotesize
{\bf Diagonal part:}
\begin{equation}
\begin{array}{l}
\tr\,{\cal C}_1{\cal C}^\dagger_1=C^3_R D_R\;, 
\nonumber \\[1.5ex]
\tr\,{\cal C}_2{\cal C}^\dagger_{2}=
\tr\,{\cal C}_3{\cal C}^\dagger_{3}=
\tr\,{\cal C}_{4}{\cal C}^\dagger_{4}=
      C_A C^2_R D_R\;, 
\nonumber \\[1.5ex]    
\tr\,{\cal C}_{5}{\cal C}^\dagger_{5}=
      C^2_A C_R D_R\;.
\nonumber \\[1.5ex] 
\end{array}\label{colo1}
\end{equation}   
\vskip 1pt
{\bf Off-diagonal part:}
\begin{equation}
\begin{array}{l}
\tr\,{\cal C}_1{\cal C}^\dagger_2=-\frac{1}{2}C_A C^2_RD_R\;, 
\nonumber \\[1.5ex]
\tr\,{\cal C}_1{\cal C}^\dagger_3=\tr\,{\cal C}_1{\cal C}^\dagger_4=
-\frac{1}{2}\left(C_R-\frac{1}{2}C_A\right)C_A C_R D_R\;,
\nonumber \\[1.5ex]
\tr\,{\cal C}_2{\cal C}^\dagger_3=\tr\,{\cal C}_2{\cal C}^\dagger_5=
-\frac{1}{4}C^2_A C_R D_R\;,
\nonumber \\[1.5ex]
\tr\,{\cal C}_3{\cal C}^\dagger_4=
-\tr\,{\cal C}_4{\cal C}^\dagger_5=
\left(C_R-\frac{1}{2}C_A\right)C_A C_R D_R\;,
\nonumber\\[1.5ex]
\tr\,{\cal C}_3{\cal C}^\dagger_5=\frac{1}{2}C^2_A C_R\;.
\nonumber\\[1.5ex]
\end{array} \label{colo2}
\end{equation} 
\vskip 1pt
{\bf Diagrams with 0:}
\begin{equation}
\begin{array}{l}
\tr\,{\cal C}_1{\cal C}^\dagger_5=\tr\,{\cal C}_2{\cal C}^\dagger_4=0\;. 
\nonumber\\[1.5ex]
\end{array}
\end{equation} 
}

With $R=C_A/C_R$, the relative conditional probability 
is  given by
\begin{eqnarray}
R^{(2)}_g(\vk \,)  &=& {\alpha_s \over \pi^2} C_R \;
\left\{ \vec{H}^2 + R \left (\vec{B}^2_1+\vec{B}^2_2+\vec{C}^2_1
+\vec{C}^2_2 \right ) + R^2 \left( \vec{B}^2_{2(12)}+ 
\vec{C}^2_{(12)} \right) \right. \nonumber \\[.5ex] 
&& -\,R\,({R\over 2})\,\vec{B}_1\cdot\vec{B}_2
\cos (t_{21}\omega_0)-R\,({R\over 2})\,\vec{C}_1\cdot\vec{B}_2
\cos (t_{21}\omega_0 +t_{10}\omega_1 )\nonumber \\[.5ex] 
&& +\,2R^2\,\vec{C}_{(12)}\cdot\vec{B}_{2(12)}
\cos (t_{10}\omega_{(12)})- R \left ( \vec{H}\cdot\vec{B}_1
\cos (t_{10}\omega_0) \right. \nonumber \\[.5ex]  
&& \left. +\,\vec{H}\cdot\vec{C}_1
\cos (t_{10}\omega_{10}) - 2\vec{C}_1\cdot\vec{B}_1
\cos (t_{10}\omega_1) \right )   \nonumber \\[.5ex] 
&& - \,R\,(1-{R\over 2})\left ( \vec{H}\cdot\vec{B}_2
\cos (t_{20}\omega_0) +\vec{H}\cdot\vec{C}_2
\cos (t_{20}\omega_{20}) \right. \nonumber \\[.5ex] 
&&\left. - \,2\vec{C}_2\cdot\vec{B}_2\cos (t_{20}\omega_2)\right ) 
 -R\,({R\over 2}) \left( \vec{C}_1\cdot\vec{B}_{2(12)}
\cos (t_{10}\omega_1-t_{21}\omega_{20}) \right. \nonumber \\[.5ex] 
&& +\,\vec{C}_1\cdot\vec{C}_{(12)}\cos (t_{10}\omega_{1(12)}
- t_{21}\omega_{20}  ) + \vec{B}_1\cdot\vec{B}_{2(12)}
\cos (t_{21}\omega_{20})  \nonumber \\[.5ex] 
&& \left.  +\,\vec{B}_1\cdot\vec{C}_{(12)}\cos (t_{10}\omega_{(12)}
+ t_{21}\omega_{20}  )\right )
 -\,R^2 \left( \vec{C}_2\cdot\vec{B}_{2(12)}
\cos (t_{10}\omega_2) \right. \nonumber \\[.5ex] 
&& +\,\vec{C}_2\cdot\vec{C}_{(12)}\cos (t_{10}\omega_{2(12)} )
 - \vec{B}_2\cdot\vec{B}_{2(12)}
\cos (t_{21}\omega_{2})  \nonumber \\[.5ex]
&& \left. \left.  -\,\vec{B}_2\cdot\vec{C}_{(12)}\cos(t_{10}\omega_{(12)}
+ t_{21}\omega_{2}  ) \right) \; \right\} \; ,
\label{neq2}
\end{eqnarray}
where we have used the concise notation introduced in Sec.~4.

\section{The $t_0 \rightarrow -\infty$ Limit in the Case of $n_s=2$}

Going back to eq.~(\ref{n2}) it is  easy to check 
the $t_0 \rightarrow -\infty$ limit  and make contact with 
BDMPS~\cite{BDMPS}. In the phases we will assume small adiabatic 
switch-off factors $\pm i\varepsilon$  when taking the corresponding limits.  

The limit of $t_0 \rightarrow -\infty$ corresponds to removing the 
interference with the jet source. In this limit only the two 
Gunion-Bertsch terms and the Gunion-Bertsch cascade term survive. 
\begin{eqnarray}
{\cal M}_2 &=& -2 i g_s \;
\vec{\epsilon}_\perp \cdot
\left\{ \vec{B}_1 e^{i t_{1} \omega_0} 
a_2 \left [ c, a_1 \right ] +  \vec{B}_2  e^{i t_{2} \omega_0} 
\left [ c, a_2 \right ] a_1 \right. \nonumber \\
\qquad && \left. + \, \vec{B}_{2(12)} 
e^{i (t_{2} \omega_{0}-t_{21}\omega_2 )} 
\left [ \left [ c, a_2 \right ] , a_1 \right ] \, \right\} \; .  
\end{eqnarray}  
This is the two-scattering generalization  of the Gunion-Bertsch  
one scattering center 
formula~\cite{GUNION} and corresponds exactly to eq.~(2.22) of 
BDMPS~\cite{BDMPS}. However, when squared there are several extra terms 
compared to the {\em one effective term} retained in  the approximations 
employed there. In particular, from eq.~(\ref{neq2})
\begin{eqnarray}
&&\lim_{t_0\rightarrow -\infty}\omega\frac{dN_g^{(2)}}{d\omega}
=\int d^2\vec{k}_\perp {dN^{(2)}_g \over dyd^2\vk_\perp} = 
{\alpha_s \over \pi^2} C_A \left \langle\int d^2 \vec{k}_\perp\;
\left\{ \vec{B}^2_1+\vec{B}^2_2 \right. \right.
\nonumber \\[.5ex] 
&&+\, R\, \vec{B}^2_{2(12)} -\,\left({R\over 2}\right)
\left( \vec{B}_1\cdot\vec{B}_2\cos (t_{21}\omega_{0})+ 
\vec{B}_1 \cdot \vec{B}_{2(12)}\cos (t_{21}\omega_{20}) \right. 
\nonumber \\[.5ex] 
&& \left. \left. \left. -\,2\, \vec{B}_2 \cdot\vec{B}_{2(12)}
\cos (t_{21}\omega_{2}) \right)\right\}\; \right\rangle
\label{n2lim} \; .
\end{eqnarray}
After taking the kinematic limits to infinity and performing the 
substitution $\vk_{\perp} =\vk^{\;'}_{\perp}-
\vq_{2\,\perp}$ in the $\vec{B}_1\cdot\vec{B}_2$ contribution one 
arrives at a result that in the large $N_c$ limit, i.e. $R=2$,  the 
weighted energy distribution of the
gluons has the form:
\begin{eqnarray}
\lim_{t_0\rightarrow -\infty}\omega\frac{dN_g^{(2)}}{d\omega} 
&=& \omega\frac{dN_g^{(2)fact.}}{d\omega}
\nonumber \\[.5ex] 
 & +&\; 3{\alpha_s \over \pi^2} C_A \left \langle\int d^2 \vec{k}_\perp\;
\left( \vec{B}_2 \cdot\vec{B}_{2(12)} \left( \cos (t_{21}\omega_{2}) -1\right)
\right)  \right \rangle
\nonumber \\[.5ex] 
 & -&\; {\alpha_s \over \pi^2} C_A \left \langle\int d^2 \vec{k}_\perp\;
\left( \vec{B}_1 \cdot\vec{B}_{2(12)} \left( \cos (t_{21}\omega_{20}) -1\right)
\right)  \right \rangle
\nonumber \\[.5ex]
&=& \omega\frac{dN_g^{(2)fact.}}{d\omega}+\omega\frac{dN_g^{(2)BDMPS}}{d\omega}
\nonumber \\[.5ex] 
 & -&\; {\alpha_s \over \pi^2} C_A \left \langle\int d^2 \vec{k}_\perp\;
\left( \vec{B}_1 \cdot\vec{B}_{2(12)} \left( \cos (t_{21}\omega_{20}) -1\right)
\right)  \right \rangle \; , \qquad
\end{eqnarray}
where direct comparison to eq.~(2.31) in Ref.~\cite{BDMPS} shows that
\begin{eqnarray}
\omega {dN^{(2)}_{BDMPS} \over d\omega} &=& 
3{C_A \alpha_s \over \pi^2} \left \langle \int d^2 \vec{k}\; 
 \left( \frac{(\vec{k}-\vec{q}_2)_\perp}{(k-q_2)_\perp^2}-
\frac{(\vec{k}-\vec{q}_1-\vec{q}_2)_\perp}
{(k-q_1-q_2)_\perp^2} \right)\cdot \right.
\nonumber \\[.5ex]
&&  \left. \cdot \left(\frac{\vec{k}_\perp}{k_\perp^{\;2}}-
\frac{(\vec{k}-\vec{q}_2)_\perp}{(k-q_2)_\perp^2}\right) 
(\cos(t_{21}\omega_2) -1) \right \rangle\nonumber \\[.5ex]
&=&3 {C_A \alpha_s \over \pi^2} \left \langle \int d^2 \vec{k}\; 
\langle \vec{B}_2\cdot\vec{B}_{2(12)}
(\cos(t_{21}\omega_2) -1) \right \rangle
\label{bdmps} \; .
\end{eqnarray}

If we also  require that $t_2 - t_1 \rightarrow \infty$ then the 
Gunion-Bertsch cascade
term  is also vanishes and we reproduce two independent 
Gunion-Bertsch scatterings with gluon radiation.

\section{Case of Three Scattering Centers}

The interaction of the incident high energy parton with 
three scattering centers accompanied by emission of one gluon
is described (in the eikonal approach) by the following fifteen diagrams:
\begin{center}
\vspace*{18.8cm}
\includegraphics{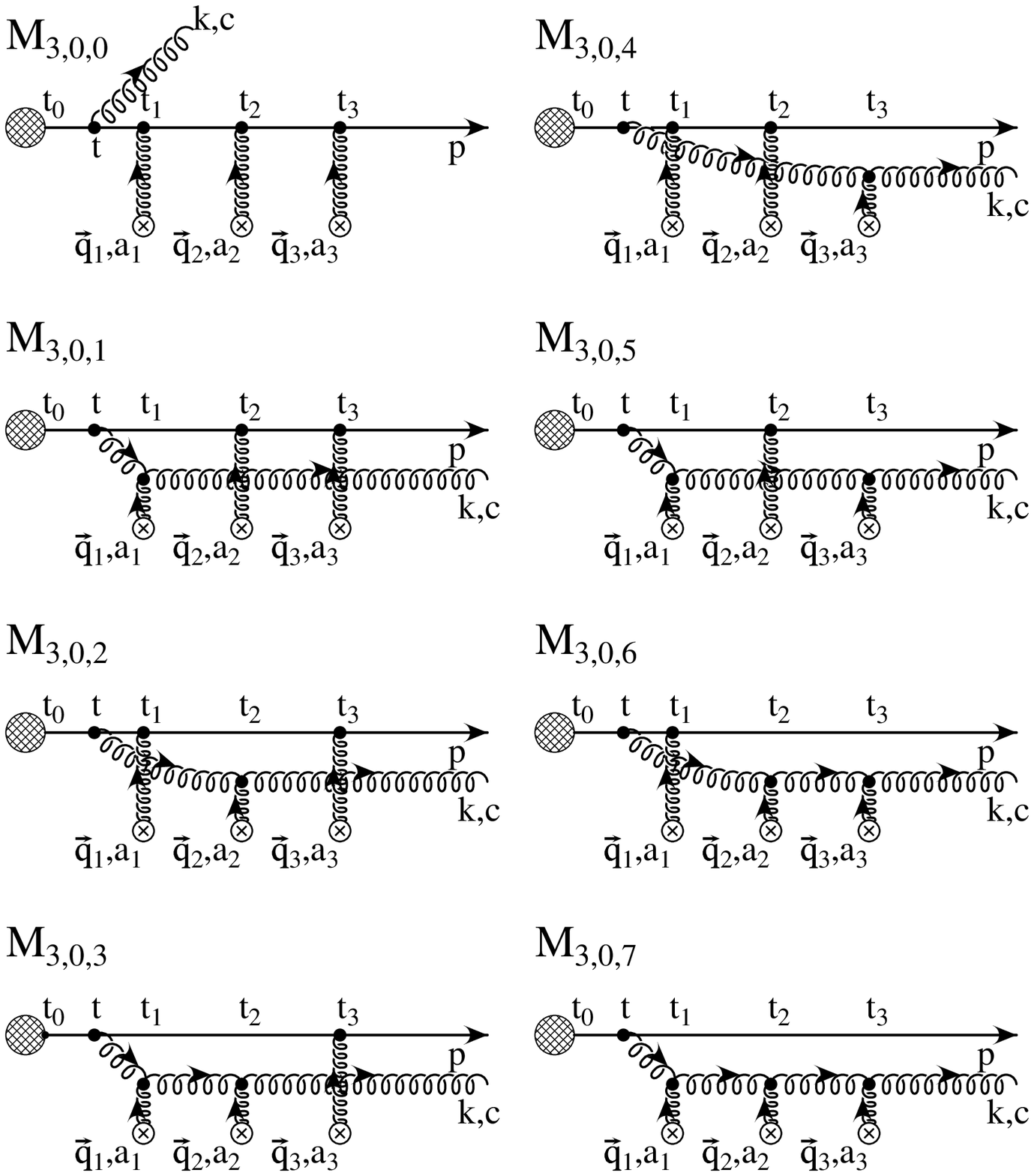}
\vskip -30pt

\newpage
\vspace*{18.5cm}
\includegraphics{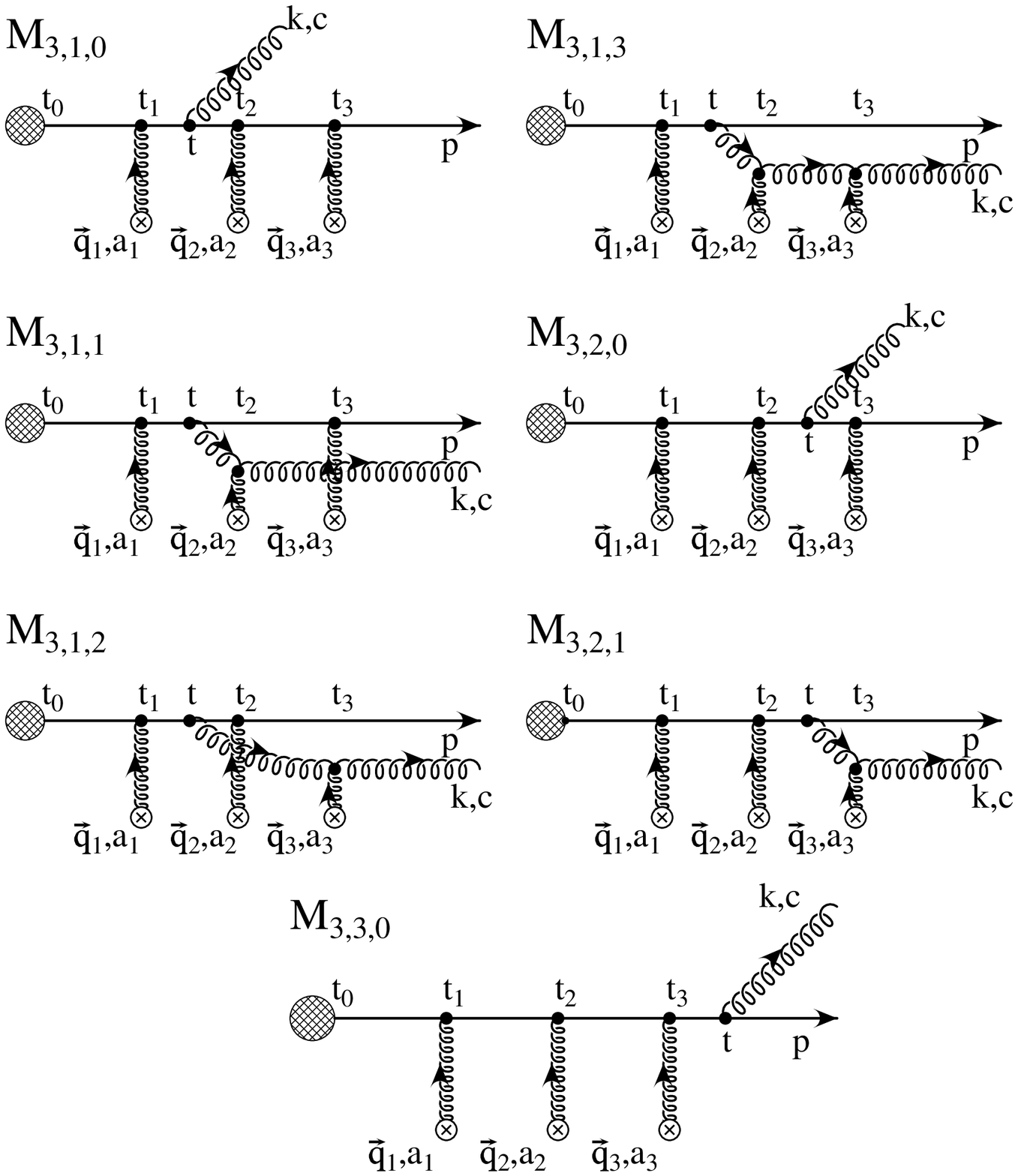}
\begin{minipage}[t]{15.0cm}
{\small {\bf Fig.~9.} The contributions to {\cal M}$_J \otimes$ {\cal M}$_3$  
soft gluon radiation amplitude in case of three scatterings.}
\end{minipage}
\end{center}
\vskip 4truemm
The corresponding matrix elements are easy to obtain through the
procedure described in Sec.~3.3 
\begin{eqnarray}
{\cal{M}}_{3,0,0} &=& 2 i g_s {\vec{\epsilon}_\perp \cdot
\vk_\perp \over k^2_\perp } 
(e^{i t_1 {k^2_\perp \over 2\omega}} 
-e^{i t_0 {k^2_\perp \over 2\omega}} )a_3 a_2 a_1 c \; , \\[.5ex]
{\cal{M}}_{3,0,1} &=& 2 i g_s {\vec{\epsilon}_\perp \cdot
(\vk - \vq_1)_\perp \over (k - q_1)^2_\perp } 
e^{i t_1 {k^2_\perp - (k - q_1)^2_\perp \over 2\omega}} \times \nonumber \\[.5ex]
 \qquad && \times (e^{i t_1 {(k - q_1)^2_\perp \over 2\omega}} 
-e^{i t_0 {(k - q_1)^2_\perp \over 2\omega}} ) 
a_3 a_2 \left [ c, a_1 \right ] \; ,  \\[.5ex]
{\cal{M}}_{3,0,2} &=& 2 i g_s {\vec{\epsilon}_\perp \cdot
(\vk - \vq_2 )_\perp \over (k - q_2)^2_\perp } 
e^{i t_2 {k^2_\perp - (k - q_2)^2_\perp \over 2\omega}} \times \nonumber \\[.5ex]
 \qquad && \times (e^{i t_1 {(k - q_2)^2_\perp \over 2\omega}} 
-e^{i t_0 {(k - q_2)^2_\perp \over 2\omega}} ) 
a_3 a_1  \left [ c, a_2 \right ] \; ,  \\[.5ex]
{\cal{M}}_{3,0,3} &=& 2 i g_s {\vec{\epsilon}_\perp \cdot
(\vk - \vq_1 -\vq_2 )_\perp \over (k - q_1 - q_2)^2_\perp } 
e^{i t_2 {k^2_\perp - (k - q_2)^2_\perp \over 2\omega}} 
e^{i t_1 {(k-q_2)^2_\perp - (k-q_1-q_2)^2_\perp \over 2\omega}} 
\times  \quad \nonumber \\[.5ex]
 \qquad && \times (e^{i t_1 {(k-q_1-q_2)^2_\perp \over 2\omega}} 
-e^{i t_0 {(k-q_1-q_2)^2_\perp \over 2\omega}} ) 
a_3\left[ \left [ c, a_2 \right ], a_1\right] \; , \\[.5ex] 
{\cal{M}}_{3,0,4} &=& 2 i g_s {\vec{\epsilon}_\perp \cdot
(\vk - \vq_3 )_\perp \over (k - q_3)^2_\perp } 
e^{i t_3 {k^2_\perp - (k - q_3)^2_\perp \over 2\omega}} \times \nonumber \\[.5ex]
 \qquad && \times (e^{i t_1 {(k - q_3)^2_\perp \over 2\omega}} 
-e^{i t_0 {(k - q_3)^2_\perp \over 2\omega}} ) 
a_2 a_1  \left [ c, a_3 \right ] \; , \\[.5ex]
{\cal{M}}_{3,0,5} &=& 2 i g_s {\vec{\epsilon}_\perp \cdot
(\vk - \vq_1 -\vq_3)_\perp \over (k - q_1 - q_3)^2_\perp } 
e^{i t_3 {k^2_\perp - (k - q_3)^2_\perp \over 2\omega}} 
e^{i t_1 {(k-q_3)^2_\perp - (k-q_1-q_3)^2_\perp \over 2\omega}} 
\times \quad \nonumber \\[.5ex]
 \qquad && \times (e^{i t_1 {(k-q_1-q_3)^2_\perp \over 2\omega}} 
-e^{i t_0 {(k-q_1-q_3)^2_\perp \over 2\omega}} ) 
a_2\left[ \left [ c, a_3 \right ], a_1\right] \; , \\[.5ex]
{\cal{M}}_{3,0,6} &=& 2 i g_s {\vec{\epsilon}_\perp \cdot
(\vk - \vq_2 -\vq_3)_\perp \over (k - q_2 - q_3)^2_\perp } 
e^{i t_3 {k^2_\perp - (k - q_3)^2_\perp \over 2\omega}} 
e^{i t_2 {(k-q_3)^2_\perp - (k-q_2-q_3)^2_\perp \over 2\omega}} 
\times  \quad \nonumber \\[.5ex]
 \qquad && \times (e^{i t_1 {(k-q_2-q_3)^2_\perp \over 2\omega}} 
-e^{i t_0 {(k-q_2-q_3)^2_\perp \over 2\omega}} ) 
a_1\left[ \left [ c, a_3 \right ], a_2\right] \; , \\[.5ex]
{\cal{M}}_{3,0,7} &=& 2 i g_s {\vec{\epsilon}_\perp \cdot
(\vk -\vq_1- \vq_2 -\vq_3 )_\perp 
\over (k -q_1- q_2 - q_3)^2_\perp } 
e^{i t_3 {k^2_\perp - (k - q_3)^2_\perp \over 2\omega}}
\times  \quad \nonumber \\[.5ex]
 \qquad && \times  
e^{i t_2 {(k-q_3)^2_\perp - (k-q_2-q_3)^2_\perp \over 2\omega}}  
e^{i t_1 {(k-q_2- q_3)^2_\perp - (k-q_1- q_2-q_3)^2_\perp 
\over 2\omega}}
\times \nonumber \quad \\[.5ex] 
 \qquad && \times  
(e^{i t_1 {(k-q_1- q_2-q_3)^2_\perp \over 2\omega}} 
-e^{i t_0 {(k-q_1- q_2-q_3)^2_\perp \over 2\omega}} ) 
\left[\left[ \left [ c, a_3 \right ], a_2\right],a_1\right] 
\; , \\[.5ex]
{\cal{M}}_{3,1,0} &=& 2 i g_s {\vec{\epsilon}_\perp \cdot
\vk_\perp \over k^2_\perp } 
(e^{i t_2 {k^2_\perp \over 2\omega}} 
-e^{i t_1 {k^2_\perp \over 2\omega}} ) a_3 a_2 c a_1 \; , \\[.5ex]
{\cal{M}}_{3,1,1} &=& 2 i g_s {\vec{\epsilon}_\perp \cdot
(\vk - \vq_2 )_\perp \over (k - q_2)^2_\perp } 
e^{i t_2 {k^2_\perp - (k - q_2)^2_\perp \over 2\omega}} \times \nonumber \\[.5ex]
 \qquad && \times (e^{i t_2 {(k - q_2)^2_\perp \over 2\omega}} 
-e^{i t_1 {(k - q_2)^2_\perp \over 2\omega}} ) 
a_3\left [ c, a_2 \right ] a_1 \; , \\[.5ex]
{\cal{M}}_{3,1,2} &=& 2 i g_s {\vec{\epsilon}_\perp \cdot
(\vk - \vq_3 )_\perp \over (k - q_3)^2_\perp } 
e^{i t_2 {k^2_\perp - (k - q_3)^2_\perp \over 2\omega}} \times \nonumber \\[.5ex]
 \qquad && \times (e^{i t_2 {(k - q_3)^2_\perp \over 2\omega}} 
-e^{i t_1 {(k - q_3)^2_\perp \over 2\omega}} ) 
a_2\left [ c, a_3 \right ] a_1 \; ,  \\[.5ex]
{\cal{M}}_{3,1,3} &=& 2 i g_s {\vec{\epsilon}_\perp \cdot
(\vk - \vq_2 -\vq_3)_\perp \over (k - q_2 - q_3)^2_\perp } 
e^{i t_3 {k^2_\perp - (k - q_3)^2_\perp \over 2\omega}} 
e^{i t_2 {(k-q_3)^2_\perp - (k-q_2-q_3)^2_\perp \over 2\omega}} 
\times \quad  \nonumber \\[.5ex]
 \qquad && \times (e^{i t_2 {(k-q_2-q_3)^2_\perp \over 2\omega}} 
-e^{i t_1 {(k-q_2-q_3)^2_\perp \over 2\omega}} ) 
\left[ \left [ c, a_3 \right ], a_2\right] a_1 \; , \\[.5ex]
{\cal{M}}_{3,2,0} &=& 2 i g_s {\vec{\epsilon}_\perp \cdot
\vk_\perp \over k^2_\perp } 
(e^{i t_3 {k^2_\perp \over 2\omega}} 
-e^{i t_2 {k^2_\perp \over 2\omega}} ) a_3 c  a_2  a_1 \; , \\[.5ex]
{\cal{M}}_{3,2,1} &=& 2 i g_s {\vec{\epsilon}_\perp \cdot
(\vk - \vq_3 )_\perp \over (k - q_3)^2_\perp } 
e^{i t_3 {k^2_\perp - (k - q_3)^2_\perp \over 2\omega}} 
\times \nonumber \\[.5ex]
 \qquad && \times (e^{i t_3 {(k - q_3)^2_\perp \over 2\omega}} 
-e^{i t_2 {(k - q_3)^2_\perp \over 2\omega}} ) 
\left [ c, a_3 \right ] a_2 a_1 \; , \\[.5ex]
{\cal{M}}_{3,3,0} &=& 2 i g_s {\vec{\epsilon}_\perp \cdot
\vk_\perp \over k^2_\perp } 
( -e^{i t_3 {k^2_\perp \over 2\omega}} ) c a_3 a_2 a_1 \; .
\end{eqnarray}

Organizing the sum  of the matrix elements in groups with common
color and phase factors we obtain
\begin{eqnarray}
{\cal M}_3 &=&  -2 i g_s  e^{i t_0 \omega_0}\times \vec{\epsilon}_\perp
\cdot  \left\{ \vec{H}a_3a_2a_1c + \vec{B}_1  e^{i t_{10} \omega_0}
a_3 a_2 \left [ c, a_1 \right ]  \right. \nonumber \\[1.5ex]
\qquad && +\, \vec{B}_2 e^{i t_{20} \omega_0}
a_3 \left [ c, a_2 \right ] a_1 + 
\vec{B}_3 e^{i t_{30} \omega_0}
\left [ c, a_3 \right ] a_2 a_1  \nonumber \\[1.5ex]
\qquad && +\,
\vec{B}_{2(12)}e^{i (t_{20} \omega_{0}-t_{21}\omega_2 )} 
a_3 \left [ \left [ c, a_2 \right ] , a_1 \right ]  
\nonumber \\[1.5ex]  \qquad && +\,
 \vec{B}_{3(13)}e^{i (t_{30} \omega_{0}-t_{31}\omega_3 )} 
a_2 \left [ \left [ c, a_3 \right ] , a_1 \right ]  
\nonumber \\[1.5ex] 
&& +\,\vec{B}_{3(23)}e^{i (t_{30} \omega_{0}-t_{32}\omega_3 )} 
 \left [ \left [ c, a_3 \right ] , a_2 \right ] a_1 
\nonumber \\[1.5ex] 
&&   +\, \vec{B}_{(23)(123)} 
 e^{-i( t_{32} \omega_{30}+t_{21}\omega_{(23)0}+t_{10}\omega_0)}
\left [ \left [ \left [ c, a_3 \right ] , a_2 \right ] , a_1 \right ]
\nonumber \\[1.5ex]
\qquad && +\, \vec{C}_1 e^{-i t_{10} \omega_{10}} 
a_3 a_2 \left [ c, a_1 \right ]  
+\vec{C}_2 e^{-i t_{20} \omega_{20}}
a_3 a_1 \left [ c, a_2 \right ]  \nonumber \\[1.5ex]
\qquad && +\, \vec{C}_3 e^{-i t_{30} \omega_{30}}
a_2 a_1 \left [ c, a_3 \right ]  
 \nonumber \\[1.5ex] 
&&+\, \vec{C}_{(12)}e^{-i (t_{10} \omega_{(12)0}
+t_{21}\omega_{20} )} 
a_3 \left [ \left [ c, a_2 \right ] , a_1 \right ]  
\nonumber \\[1.5ex] 
&&+ \, \vec{C}_{(13)}e^{-i (t_{10} \omega_{(13)0}
+t_{31}\omega_{30} )} 
a_2 \left [ \left [ c, a_3 \right ] , a_1 \right ]  
\nonumber \\[1.5ex]
&&+\, \vec{C}_{(23)}e^{-i (t_{20} \omega_{(23)0}
+t_{32}\omega_{30} )} 
a_1 \left [ \left [ c, a_3 \right ] , a_2 \right ]   
\nonumber \\[1.5ex]
&&  \left. +\, \vec{C}_{(123)} 
 e^{-i( t_{32} \omega_{30}+t_{21}\omega_{(23)0}-t_{10}\omega_{(123)0})}
\left [ \left [ \left [ c, a_3 \right ] , a_2 \right ] , a_1 \right ]
\; \right\} \; . 
\label{n3} \end{eqnarray}

Eleven distinct color factors are present for $n_s=3$. We label them as
follows:
\begin{equation}
    \begin{array}{lll}
     \quad {\cal C}_1=a_3 a_2a_1c \; , &
     \quad {\cal C}_2= a_3a_2 \left [c ,a_1\right] \; , & 
     \quad {\cal C}_3= a_3 \left [ c,a_2 \right] a_1 \; , 
     \nonumber \\[1.5ex]
     \quad {\cal C}_4= \left [c ,a_3 \right] a_2 a_1 \; ,&
     \quad {\cal C}_5=a_3 a_1 \left [c,a_2\right ] \; ,  &
     \quad {\cal C}_6=a_2 a_1 \left [c,a_3\right ] \; ,
     \nonumber \\[1.5ex]
     \quad {\cal C}_7 =a_3 \left [\left [c,a_2 \right],a_1 \right ] 
     \; , &
     \quad {\cal C}_8=a_2 \left [\left[c,a_3\right ],a_1\right ] 
     \; , &
     \quad {\cal C}_9= \left [\left[c,a_3\right]a_2\right]a_1 
     \; ,  
     \nonumber \\[1.5ex]
     \quad {\cal C}_{10}=a_1 \left [\left[c ,a_3\right ],a_2 \right] 
     \; , & 
     \quad {\cal C}_{11}=\left [ \left [\left [c,a_3\right]a_2\right ] ,
     a_1 \right] \; . 
 \end{array}  
\end{equation}

Computing the quantum amplitude requires knowing the color traces.
These traces can be organized into a symmetric matrix:
\vskip 9pt
{\footnotesize
{\bf Diagonal terms:}
\begin{equation}
\begin{array}{l}
\tr\,{\cal C}_1{\cal C}^\dagger_1=C^4_R D_R\;, 
\nonumber \\[1.5ex]
\tr\,{\cal C}_2{\cal C}^\dagger_{2}=
\tr\,{\cal C}_3{\cal C}^\dagger_{3}=
\tr\,{\cal C}_{4}{\cal C}^\dagger_{4}=
\tr\,{\cal C}_{5}{\cal C}^\dagger_{5}=
\tr\,{\cal C}_{6}{\cal C}^\dagger_{6}=
      C_A C^3_R D_R\;,  \nonumber \\[1.5ex]    
\tr\,{\cal C}_{7}{\cal C}^\dagger_{7}=
\tr\,{\cal C}_{8}{\cal C}^\dagger_{8}=
\tr\,{\cal C}_{9}{\cal C}^\dagger_{9}=
\tr\,{\cal C}_{{10}}{\cal C}^\dagger_{{10}}=
      C^2_A C^2_R D_R\;,
\nonumber \\[1.5ex] 
\tr\,{\cal C}_{{11}}{\cal C}^\dagger_{{11}}=C^3_A C_R D_R\;. 
\nonumber \\[1.5ex]      
\end{array}\label{diag3}
\end{equation}   
\vskip 1pt
{\bf Diagrams with 0:}
\begin{equation}
\begin{array}{l}
\tr\,{\cal C}_1{\cal C}^\dagger_7=
\tr\,{\cal C}_1{\cal C}^\dagger_8=
\tr\,{\cal C}_1{\cal C}^\dagger_9=
\tr\,{\cal C}_2{\cal C}^\dagger_5=
\tr\,{\cal C}_2{\cal C}^\dagger_6=
\tr\,{\cal C}_2{\cal C}^\dagger_{9}=
0\;, 
\nonumber\\[1.5ex]
\tr\,{\cal C}_3{\cal C}^\dagger_6=
\tr\,{\cal C}_3{\cal C}^\dagger_{8}=
\tr\,{\cal C}_5{\cal C}^\dagger_{11}=
\tr\,{\cal C}_6{\cal C}^\dagger_{11}=
\tr\,{\cal C}_7{\cal C}^\dagger_{10}=
\tr\,{\cal C}_8{\cal C}^\dagger_{10}=
0\;. 
\nonumber\\[1.5ex]
\end{array}
\end{equation} 
\vskip 1pt
{\bf Traces involving the ``Wheel'' diagram} (see Appendix D):
\begin{equation}
\begin{array}{l}
\tr\,{\cal C}_1{\cal C}^\dagger_{10}=
\tr\,{\cal C}_2{\cal C}^\dagger_{11}=
\tr\,{\cal C}_5{\cal C}^\dagger_{6}=
\tr\,{\cal C}_7{\cal C}^\dagger_{8}=
     W - \frac{1}{8} C^3_A C_R D_R\;, 
\nonumber \\[1.5ex]
\tr\,{\cal C}_1{\cal C}^\dagger_{11}=
\tr\,{\cal C}_2{\cal C}^\dagger_{10}=
\tr\,{\cal C}_5{\cal C}^\dagger_{8}=
\tr\,{\cal C}_6{\cal C}^\dagger_{7}=
     -W + \frac{1}{8} C^3_A C_R D_R\;.
\nonumber \\[1.5ex]
\end{array}\label{Wfact}
\end{equation} 
\vskip 1pt
{\bf Remaining off-diagonal terms:}
\begin{equation}
\begin{array}{l}
\tr\,{\cal C}_1{\cal C}^\dagger_2=-\frac{1}{2}C_A C^3_R D_R\;,
\nonumber \\[1.5ex] 
\tr\,{\cal C}_1{\cal C}^\dagger_{3}=
\tr\,{\cal C}_1{\cal C}^\dagger_{5}=
     -\frac{1}{2} \left(C_R-\frac{1}{2}C_A\right) C_A C^2_R D_R\;,
\nonumber \\[1.5ex]
\tr\,{\cal C}_1{\cal C}^\dagger_{4}=
\tr\,{\cal C}_1{\cal C}^\dagger_{6}=
     -\frac{1}{2} \left(C_R-\frac{1}{2}C_A\right)^2 C_A C_R D_R\;, 
\nonumber \\[1.5ex]
\tr\,{\cal C}_2{\cal C}^\dagger_{3}=
\tr\,{\cal C}_2{\cal C}^\dagger_{7}=
\tr\,{\cal C}_3{\cal C}^\dagger_{4}=
\tr\,{\cal C}_3{\cal C}^\dagger_{9}=
     -\frac{1}{4} C^2_A C_R^2 D_R\;, 
\nonumber \\[1.5ex]
\tr\,{\cal C}_2{\cal C}^\dagger_{4}=
\tr\,{\cal C}_2{\cal C}^\dagger_{8}=
\tr\,{\cal C}_3{\cal C}^\dagger_{10}=
\tr\,{\cal C}_4{\cal C}^\dagger_{5}=
\tr\,{\cal C}_5{\cal C}^\dagger_{9}=
\tr\,{\cal C}_5{\cal C}^\dagger_{10}=
     -\frac{1}{4} \left(C_R-\frac{1}{2}C_A\right) C^2_A C_R D_R\;, 
\nonumber \\[1.5ex]
\tr\,{\cal C}_3{\cal C}^\dagger_{5}=
     \left(C_R-\frac{1}{2}C_A\right) C_A C^2_R D_R\;, 
\nonumber \\[1.5ex]
 \tr\,{\cal C}_3{\cal C}^\dagger_{7}=
 \tr\,{\cal C}_4{\cal C}^\dagger_{9}=
-\tr\,{\cal C}_5{\cal C}^\dagger_{7}=
-\tr\,{\cal C}_6{\cal C}^\dagger_{8}=
   \frac{1}{2} C^2_A C^2_R D_R\;, 
\nonumber \\[1.5ex]
\tr\,{\cal C}_3{\cal C}^\dagger_{11}=
\tr\,{\cal C}_4{\cal C}^\dagger_{7}=
\tr\,{\cal C}_7{\cal C}^\dagger_{9}=
\tr\,{\cal C}_7{\cal C}^\dagger_{11}=
     -\frac{1}{8} C^3_A C_R D_R\;, 
\nonumber \\[1.5ex]
\tr\,{\cal C}_4{\cal C}^\dagger_{6}=
     \left(C_R-\frac{1}{2}C_A\right)^2 C_A C_R D_R\;, 
\nonumber \\[1.5ex]
 \tr\,{\cal C}_4{\cal C}^\dagger_{8}=
 \tr\,{\cal C}_4{\cal C}^\dagger_{10}=
-\tr\,{\cal C}_{6}{\cal C}^\dagger_{9}=
-\tr\,{\cal C}_{6}{\cal C}^\dagger_{10}=
     \frac{1}{2} \left(C_R-\frac{1}{2}C_A\right) C^2_A C_R D_R\;, 
\nonumber \\[1.5ex]
\tr\,{\cal C}_4{\cal C}^\dagger_{11}=
-\tr\,{\cal C}_8{\cal C}^\dagger_{9}=
-\tr\,{\cal C}_8{\cal C}^\dagger_{11}=
     \frac{1}{4} C^3_A C_R D_R\;, 
\nonumber \\[1.5ex]
\tr\,{\cal C}_9{\cal C}^\dagger_{10}=
     \left(C_R-\frac{1}{2}C_A\right) C^2_A C_R D_R\;, 
\nonumber \\[1.5ex]
 \tr\,{\cal C}_9{\cal C}^\dagger_{11}=
-\tr\,{\cal C}_{10}{\cal C}^\dagger_{11}=
     \frac{1}{2} C^3_A C_R D_R\;. 
\nonumber \\[1.5ex]
\end{array} \label{diag31} 
\end{equation} 
}

\newpage

\section{The Color ``Wheel'' Diagrams}
In the case of three scattering centers some nontrivial 
color factors appear for the first time  in eq.~(\ref{Wfact}). 
They correspond  diagrammatically to the ``wheel''
diagrams  shown in Fig.~10. 

\begin{center}
\vspace*{8.1cm}
\includegraphics{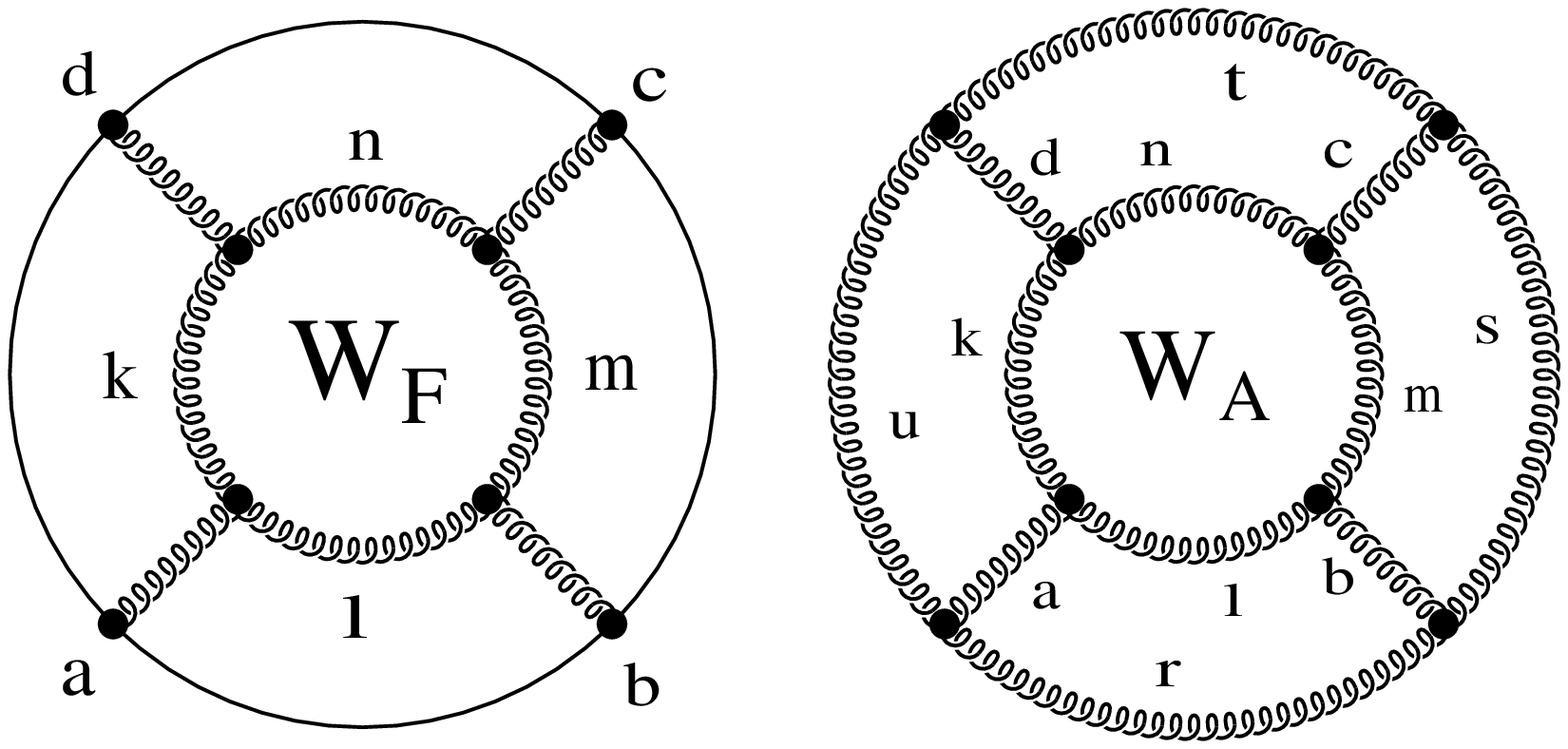}
\vskip -55pt
\begin{minipage}[t]{15.0cm}
{\small {\bf Fig.~10.}
The ``wheel'' diagram in the fundamental and the adjoint representations.} 
\end{minipage}
\end{center}
\vskip 4truemm

For the fundamental representation
\begin{eqnarray}
\left \{T_a,T_b   \right\}&=&\frac{1}{N}\delta_{ab}+d_{abc} T_c\;,
\end{eqnarray} 
It is useful to introduce
the complex tensor
\begin{equation}
h_{abc}=d_{abc}+i f_{abc}\;, \quad h_{abc}=h_{bca}=h_{cab}\;,
\quad h_{aab}=0\;. 
\end{equation}
If the physical case of  $SU(3)$,  we find 
\begin{eqnarray}
W_F & =& \tr \left( T_a T_b T_c T_d \right )
f_{kal} f_{lbm} f_{mcn}  f_{ndk}
=\left (\frac{1}{12}\delta_{ab}\delta_{cd}+
\frac{1}{8}h_{abr}h_{rcd} \right ) \times
\nonumber \\[1.5ex]
&& \times \frac{1}{4} \left (5\delta_{ab}\delta_{cd}+
\delta_{ac}\delta_{bd}+5\delta_{ad}\delta_{bc}-
6d_{acr}d_{rbd} \right )=\frac{33}{2}
\end{eqnarray} 
for the fundamental representation and
\begin{eqnarray}
W_A & =& (f_{kal} f_{lbm} f_{mcn}  f_{ndk} )
(f_{rau} f_{udt} f_{tcs}  f_{sbr} ) =
\nonumber \\[1.5ex]
&=&\frac{1}{4} \left (5\delta_{ab}\delta_{cd}+
\delta_{ac}\delta_{bd}+5\delta_{ad}\delta_{bc}-
6d_{acr}d_{rbd} \right )\times 
\nonumber \\[1.5ex]
&&\times \frac{1}{4} \left (5\delta_{ad}\delta_{cb}+
\delta_{ac}\delta_{bd}+5\delta_{ab}\delta_{dc}-
6d_{acr}d_{rdb} \right ) = 189\;.
\end{eqnarray} 
for the adjoint representation. 
\vskip 13pt
{\hspace*{1.6cm} 
\begin{tabular}{|c||c|c|c|c|} \hline 
&&&& \\
$N_c$ &   2 & 3 & 4 &  5  \\ &&&& \\ \hline
&&&& \\
$W_F$ &  $\frac{9}{4} $ &   $\frac{33}{2}$ & $\frac{135}{2}$ & $\frac{405}{2}$
    \\ &&&& \\ \hline
&&&& \\
\ \ \ $W_A$ \ \ \ & \ \ \  \ 24\ \ \ \  & \ \ \ 189 \ \ \ & \ \ \ 840 \ \ \ & 
\ \ 2775 \ \   \\ &&&& \\ \hline
\end{tabular} }
\vskip 20pt
\begin{minipage}[t]{13.7cm}
{\small {\bf Table~1.}
The $N_c$ dependence of the wheel diagram to be found
numerically. }
\end{minipage}

\end{appendix}

\vskip 30pt

\hrule


\begin{thebibliography}{80}


\bibitem{BJ}J.D. Bjorken, Fermilab-Pub-82/59-THY (1982) and
        erratum (unpublished).
\bibitem{THOMA}M.~H.~Thoma and M.~Gyulassy, Nucl. Phys. B {\bf 351}
        (1991) 491. 
\bibitem{MGMP}M. Gyulassy and M. Pl\"umer, Phys. Lett. B {\bf 243}
        (1990) 432. 
\bibitem{GPTW}M. Gyulassy, M. Pl\"umer, M.H. Thoma and X.-N. Wang,
        Nucl. Phys. A {\bf 538} (1992) 37c.
\bibitem{MGXW92}X.-N. Wang and M.~Gyulassy, Phys. Rev. Lett.
        {\bf 68} (1992) 1480.
\bibitem{GUNION}J.F. Gunion and G. Bertsch, Phys. Rev. D {\bf 25}
        (1982) 746.
\bibitem{BRODS}S.J. Brodsky and P. Hoyer, Phys. Lett. B {\bf 298}
        (1993) 165.
\bibitem{MGXW}M.~Gyulassy and X.-N.~Wang, Nucl. Phys. B {\bf 420}
        (1994) 583.
\bibitem{BDMPS} R. Baier, Yu.L. Dokshitzer, A.H. Mueller, S. Peign\'e,
         D. Schiff,  Nucl. Phys. B {\bf 483} (1997) 291.
\bibitem{LPM}L.D. Landau and I.J. Pomeranchuk, Dolk. Akad. Nauk. 
         SSSR {\bf 92} (1953) 92; A.B. Migdal, Phys. Rev. {\bf 103}
         (1956) 1811; E.L. Feinberg and I.J. Pomeranchuk, 
         Suppl. Nuovo. Cimento {\bf 3} (1956) 652; 
         Phys. JETP {\bf 23} (1966) 132.
\bibitem{Dong}  D.W. Dong and M. Gyulassy, Phys. Rev. E {\bf 47} 
         (1993) 2913.
\bibitem{GLVII} M. Gyulassy, P. L\'evai, and I. Vitev,
            in preparation;  preliminary numerical results presented
            at RHIC Winter Workshop, Jan 7-9, 1999, LBNL, see \\
            $http://sseos.lbl.gov/~nxu/workshop/talk06/ppframe.html$
\bibitem{hijing}X.-N. Wang and M. Gyulassy, Phys. Rev. D {\bf 44} (1991)
         3501; X.-N. Wang, Comp. Phys. Comm. {\bf 83} (1994) 307.
\bibitem{JETSET}
         T.~Sj\"{o}strand, Comput. Phys. Commun. {\bf 27} (1982) 243.
\bibitem{WiedGY}U.A. Wiedemann and M. Gyulassy,
                hep-ph/9906257, submitted to NPB.
\bibitem{Zakharov} B.G. Zhakharov, JETP Letters {\bf 63} (1996) 952; 
         {\it ibid.} {\bf 65} (1997) 615.
\bibitem{BDMS} R. Baier, Yu.L. Dokshitzer, A.H. Mueller,  S. Peign\'e, 
          D. Schiff, Nucl. Phys. B {\bf 484} (1997) 265; 
          Nucl. Phys. B {\bf 531} (1998) 403;
           Phys. Rev. C {\bf 58} (1998) 1706. 
\bibitem{GLVQM99} M. Gyulassy, P. L\'evai, I. Vitev,
     {\it Jet quenching in thin plasmas},
     hep-ph/9907343, to appear in Nucl. Phys. A.
\bibitem{Field} R. Field, {\it Applications of Perturbative QCD},
         Addison-Wesley, 1989.
\bibitem{PYTHIA}T. Sj\"{o}strand and M. van Zijl, Phys. Rev. D 
           {\bf 36} (1987) 2019; 
         T. Sj\"{o}strand, Comp. Phys. Commun. {\bf 39} (1986) 347;
         T. Sj\"{o}strand and M. Bengtsson, {\em ibid.} {\bf 43} (1987) 367.
\bibitem{HIJINGPP}X.-N. Wang and M. Gyulassy, Phys. Rev. D {\bf 45}
        (1992) 844.
\bibitem{Blaizot}J.P. Blaizot and L. McLerran, Phys. Rev. D {\bf 34} (1986) 2739.
\bibitem{Uli}M. Rammerstorfer, U. Heinz, Phys. Rev. D {\bf 41} (1990) 306.
\bibitem{ZPC}M. Gyulassy, Y. Pang, and B. Zhang, Nucl. Phys. A {\bf 626} (1997) 999.
\bibitem{Cvit}  P. Cvitanovic,  Phys. Rev. D {\bf 14} (1976) 1536.
\end{thebibliography}
\end{document}